\documentclass[oneside,reqno]{amsart}

\usepackage[margin=3.5cm]{geometry}

\usepackage{rotating}

\usepackage{setspace}

\usepackage{pifont}

\usepackage{graphicx} 
\usepackage{array} 
\usepackage{bm}
\usepackage{ctable} 
\usepackage{subfig} 
\usepackage{caption} 
\usepackage{footnote}
\usepackage{threeparttable}
\usepackage{color}
\usepackage{rotating}
\usepackage{lscape}
\usepackage{tensor}
\usepackage[linewidth=1.5pt]{mdframed}
\usepackage{hyperref} 

\usepackage[round]{natbib}
\makeatletter 
\renewcommand\@biblabel[1]{#1.} 
\makeatother %

\graphicspath{{Figures/}} 

\begin{document}

\title{Flexible behavioral capture-recapture modelling}

\author[D. Alunni Fegatelli]{Danilo Alunni Fegatelli}
\address{Dipartimento di Sanit\`a  Pubblica e Malattie Infettive\\
 Sapienza Universit\`a di Roma\\
 Piazzale A. Moro 5\\00185~Roma\\Italy}
\email{danilo.alunnifegatelli@uniroma1.it}
\author[L.   Tardella]{Luca Tardella}
\address
{Dipartimento di Scienze statistiche\\
 Sapienza Universit\`a di Roma\\
 Piazzale A. Moro 5\\00185~Roma\\Italy}
\email{luca.tardella@uniroma1.it}

\thanks
{Version \today} 

\begin{abstract}
We develop some new strategies for building and fitting new flexible classes
of parametric capture-recapture models for closed populations which 
can be used to address a better understanding of  behavioural
patterns. We first rely on a conditional
probability parameterization and review how to regard a large subset of 
standard capture-recapture models as a
suitable partitioning in equivalence classes of the full set of
conditional probability parameters. We then 
propose 
the use of new suitable quantifications 
of the conditioning binary partial capture histories as a device for 
enlarging the scope of flexible  behavioural models
and also 
exploring the range of all possible partitions.  
We show how one can easily find unconditional MLE of such models 
within a generalized linear model framework.
We illustrate the potential of our approach
with the analysis of some known datasets and a simulation study. 
\\
\end{abstract}



\maketitle

\makeatletter \@setaddresses \makeatother \renewcommand{\addresses}{}

\newpage

\section{Introduction}
\label{sec:intro}

Multiple capture-recapture models are successfully employed to
infer the unknown size and characteristics of a finite population whose complete
enumeration is difficult or impossible due to the elusive nature of its 
units. These models are also routinely used in fields different from ecology
such as software engineering,  social sciences and epidemiology. 
Much progress has been made by researchers to 
enlarge the scope of available models and refine inferential
techniques.
There are now many available monographies and review articles 
which can offer a wide perspective of the current state of the art 
\citep{
whit:ande:burn:otis:1982,sebe:1987,schw:sebe:1999,borc:buck:zucc:2002,
amstrup:2005,bohn:2008,morg:mccr:2013,royl:cand:soll:gard:2013}. 
In this paper we are interested in developing tools for a better
understanding of the behavioural response to capture within 
a suitable and general model framework. Indeed 
empirical
studies have provided evidence that mice, voles, small
mammals and butterflies, among others, often exhibit a response
to capture \citep{Yang:Chao:2005,rams:seve:2010}. 
However, relevant response to capture 
is also at stake in studies involving human population 
\citep{Farc:Scac:2014}. 
The most classical and basic way to account for behavioural response 
is to assume that once a unit/animal is captured its probability of
being recaptured in all future trapping occasions is modified 
permanently. This enduring effect
is
called trap-happiness or trap-shyness 
effect
according to whether the recapture probability becomes larger or smaller. 
This very simple one-parameter model flexibility sheds some light on the population
under study and the presence of a 
behavioral effect can have a great impact on the estimate of the 
unknown population size 
\citep{Yip:Xi:Chao:Hwan:esti:2000,Hwan:Chao:Yip:cont:2002,hwang:hugg:2011, 
Lee:Chen:baye:1998, 
Chao:Chu:Hsu:capt:2000,Lee:Hwan:Huan:baye:2003,Ghos:Norr:baye:2005,aft:2012}. 
However this specific type of behavioural effect is certainly only a
limited  device to
approach the understanding of complex behavioral patterns that can be originated
in multiple capture-recapture designs. 
In fact, an extended, more flexible perspective has been introduced in 
\cite{Yang:Chao:2005} where an ephemeral behavioural effect is
modelled by using 
a Markov chain of the sequence of consecutive captures and 
\cite{Bart:Penn:clas:2007} developed a more complex model framework to 
account for dependence and heterogeneity with a hidden Markov model
for the sequence of capture probabilities. More recently new ideas
have been put forward by \cite{rams:seve:2010}
and \cite{Farc:2011} to enlarge the scope of possible behavioural
patterns with 
new instances  of enduring and ephemeral behavioural
effects. In order to provide a general and flexible framework to deal with
behavioural effects we start form the same idea in 
\cite{Farc:2011}
to fully parameterize the 
joint probabilities of the observable capture histories
in terms of conditional probabilities
and we show how the introduction of suitable behavioural
covariates can help understanding and fitting meaningful behavioural models.
We show how the appropriate handling of these covariates employed within a
generalized linear model framework can help the researcher to improve 
model fitting of capture-recapture experiments with new interpretable
behavioral patterns. 
Differently from the aforementioned articles we privilege the use of
the unconditional likelihood for making inference. 
The paper is organized as follows: in Section~\ref{sec:setup} we 
introduce basic notation for our set up, the saturated parameterization for
the probability of observing a sequence of consecutive binary outcomes
corresponding to all capture occasions and the subset of all possible 
reduced models. In Section~\ref{sec:covariate} we explain how one can
define a time-dependent {\em behavioural covariate} to be
exploited in a generalized linear model framework in order to achieve
the two-fold goal of {\em i)} enlarging the scope of available behavioural effect
models and {\em ii)}  recover many (if not all) the existing ones by  
appropriate partitioning of the range of the behavioral covariate. 
In particular we show how a specific instance of numerical covariate
can be obtained as a quantification of the binary subsequences of partial
capture histories and can be thought of as a memory-effect
covariate. In fact we will show the relation between this covariate
and the class of Markovian models of arbitrary order.  
In Section~\ref{sec:inference}
we explain how one can infer on unknown parameters 
through the maximization of unconditional likelihood  
and how 
easily it can be
implemented recycling standard generalized linear model routines.
In Section~\ref{sec:examples} 
we show the usefulness of our covariate approach in discovering 
better parsimonious models with some real data examples. 
In Section~\ref{sec:simulation} we verify the ability of model selection
criterion to identify and distinguish among different behavioural
patterns with a simulation study.  Section~\ref{sec:close}  
closes with some remarks and a discussion on future developments.

\section{Saturated and reduced models based on partitions of conditional probabilities}
\label{sec:setup}

Let us consider a discrete-time closed capture-recapture experiment 
in which the unknown population size $N$ is assumed to be constant 
and individual trappings are recorded in $t$ consecutive times. 
Moreover, we suppose that all units act independently, there is no misclassification 
i.e. all individuals are always recorded correctly and do not lose
their marks. 
For notational convenience one can assume
that units captured during the study are labelled from $1$ to $M$ 
and those not captured from $M+1$ to $N$. 
It is clear that we can observe only the firsts $M$ rows of the matrix 
$\textbf{X}$ with generic entri $x_{ij}$ with $i=1,...,N$ and $j=1,...,t$. 
Denoting with $\mathcal{X}=\{0,1\}$, the space of all possible capture histories 
for each unit is $\mathcal{X}^{t}=\lbrace 0,1\rbrace^{t}$ 
while the set of all observable capture histories is 
$\mathcal{X}^{t}_{*}=\mathcal{X}^{t}\setminus (0,\dots ,0)$ 
since the unobserved units are not sampled.
As a starting point no individual heterogeneity is assumed
for the probability of being captured at each time. 
We will discuss later on relaxation of  this assumption.

In order to setup a natural 
flexible framework 
for characterizing the 
fundamental set of 
probabilities of all possible {\em complete} capture histories
we follow 
\cite{Farc:2011} and  
we rely upon 
the capture probabilities conditioned on each possible 
{\em partial} capture history as follows
\begin{eqnarray*} 
\begin{cases}
p_{1}()=Pr(X_{i1} = 1)  \\
p_{j}(x_{i1},...,x_{ij-1})=Pr(X_{ij}=1|x_{i1},...,x_{ij-1} ) \qquad \forall j>1 \: ,\: \forall (x_{i1},\dots ,x_{ij-1})\in \mathcal{X}^{j-1}
\end{cases}  
\end{eqnarray*}
All these conditional probabilities can be arranged with a
natural/conventional order in a $2^{t}-1$ dimensional vector denoted with
$\textbf{p}=$ $(p_{1}()$, $p_{2}(0)$, $p_{2}(1)$, $p_{3}(0,0)$, $p_{3}(0,1)$, $p_{3}(1,0)$, $...$, $p_{t}(0,...,0)$, $...$, $p_{t}(1,...,1))$
where, for example, the element $p_{3}(0,1)$ represents the
probability of being captured at time 3 given that the unit is not
captured in the first occasion while it is captured in the second
occasion. 
For notational convenience and space saving we will often remove
commas between binary digits when representing partial capture
histories.  
The initial empty brackets $()$ are understood as the absence of previous capture history at time 1. The vector $\textbf{p}$ can be seen as a convenient reparameterization of 
the joint probabilities 
corresponding to all $2^{t}$ complete capture history configurations in $\mathcal{X}^{t}$. 
The conditional probabilities, rather than the joint probabilities, are more easily 
interpreted  in the process of modelling the consequences 
determined by the change of behaviour due to a particular previous trapping history. \\
Notice that 
under the saturated reparameterization the probability of never being observed during the experiment is
\begin{eqnarray}
\label{P0}
P_{0}=\left[ (1-p_{1}())\prod_{j=2}^{t}(1-p_{j}(0,\dots ,0))\right]
\end{eqnarray}
This is one of the fundamental quantities for the estimation of 
the parameter of interest via likelihood maximization since 
$\hat{N} = \frac{M}{1-\hat{P}_0}$
where $\hat{P}_0$ is the MLE of $P_0$.
This is of course true by definition of the conditional likelihood
approach but it is still true with the unconditional likelihood
provided that $\hat{P}_0$ is jointly determined with $\hat{N}$
according to the definition of the unconditional (or complete) 
likelihood.

From the saturated parametrization 
based on $\mathbf{p}$
one can specify a parsimonious
nested model based on a suitable partition of the conditional
probabilities in $\textbf{p}$ in terms of equivalence classes. Let $H$ 
be the set of all partial capture histories: $H=\left\lbrace\right.$
() , (0), (1), (00), (10), (01), (11), $ \left. \dots \right\rbrace
=\cup_{j=0}^{t-1}\mathcal{X}^{j}$ where $\mathcal{X}^{0}=\left\lbrace
  () \right\rbrace $. Let us denote with  $\mathcal{H}_{B}$ one of the possible partitions of $H$ in $B$ disjoint subsets
\begin{eqnarray*}
\mathcal{H}_{B}=\left\lbrace H_{1},\dots , H_{b}, \dots , H_{B} \right\rbrace
\end{eqnarray*}
where each $H_b \subset H$. The role of the index set $H$ is to 
list all the partial capture histories which 
may yield possible changes in the 
conditional capture probability depending on the past. 
Let us denote a generic partial capture history as follows
$\mathbf{x}=(x_{1},\dots , x_{l_{\mathbf{x}}})$
where $l_{\mathbf{x}}$ is the length of the binary vector.
For each partition $\mathcal{H}_{B}$
we consider a corresponding reduced parameter vector of 
probabilities denoted with 
$\mathbf{p}_{\mathcal{H}_{B}}=(p_{H_{1}},\dots ,p_{H_{B}})$.
The partition of capture histories in equivalence classes is such that 
\begin{eqnarray*}
\forall \quad \mathbf{x},\mathbf{x}' \in H_{b} \quad \Rightarrow \quad p_{l_{\mathbf{x}}+1}(\mathbf{x})=p_{l_{\mathbf{x}'}+1}(\mathbf{x}')=p_{H_{b}} &\qquad \forall b=1,\dots , B 
\end{eqnarray*}
Notice that
when there is absence of previous capture history ($\mathbf{x}=()$) we
have $l_{\mathbf{x}}=0$. \\
With the partition ${\mathcal H}_B$ of subsets of $H$
representing equivalence classes 
we have just reviewed under a more convenient formalization the linear
constrained approach in \cite{Farc:2011} which indeed can be seen as
stemmed from the works of \cite{Hugg:on:1989} and  \cite{Alho:1990}.

As a simple example of our formalization based on partitions of
subsets of $H$
as opposed to the linear constrained approach 
one can consider model $M_{b}$. 
Indeed it can be defined using two 
blocks of equality constraints
\begin{eqnarray*}
M_{b} : 
\begin{cases}
p_{1}()=p_{2}(0)=p_{3}(0,0)=\cdots =p_{t}(0,\dots ,0) = \pi_f\\
p_{2}(1)=p_{3}(1,0)=p_{3}(0,1)=\cdots =p_{t}(1,\dots ,1) = \pi_r
\end{cases}
\end{eqnarray*}
It is easy to verify that if we interpret $\pi_f$ as the probability of
first capture and $\pi_r$ as the probability of  being recaptured 
one gets
the most simple form of behavioural model with enduring effect
usually denoted with $M_{b}$.  
Equivalently, in our partition notation,  
model $M_{b}$ corresponds to a bipartition $\mathcal{H}_{2}(M_{b})=\left\lbrace H_{1}, H_{2}\right\rbrace $ such that
\begin{eqnarray}
\label{Mb-part}
\mathcal{H}_{2}(M_{b})=
\begin{cases}
H_{1}=\left\lbrace (), (0), (00), \dots ,(0\dots 0)\right\rbrace =
\mathcal{X}^{0} \cup \left\lbrace \mathbf{x} \in \cup_{j=1}^{t-1}\mathcal{X}^{j} : \sum_{j=1}^{l_{\mathbf{x}}} x_{j} =0 \right\rbrace \\
H_{2}=H\setminus H_{1} 
\end{cases}	
\end{eqnarray}
and the vector of parameters $(\pi_f,\pi_r)$ is represented 
in our notation 
as $\mathbf{p}_{\mathcal{H}_{2}(M_{b}) } = (p_{H_{1}},p_{H_{2}})$.

In \cite{Farc:2011} is also shown that many other models proposed in the
literature 
such as model $M_{0}$,  $M_{c_{k}}$, $M_{c_{k}b}$, $M_{t}$ 
can be recovered as special cases of model with saturated parameterization 
$\textbf{p}$
subject to specific linear constraints.
In the following we prefer to index parameters with the partition notation
and we refer to the reduced parametrization with the symbol
$\mathbf{p}_{\mathcal{H}_{B}}=(p_{H_{1}},\dots ,p_{H_{B}})$ 
corresponding to the uniquely identified  
conditional probabilities associated to 
the partition $\mathcal{H}_{B}$.

We now briefly provide details on two other meaningful partitions corresponding to 
Markovian models of order 1 and 2 respectively.
They will be used 
and clarified more extensively in the next section. 
In the generic model $M_{c_{k}}$, for each unit, capture probability
at some stage $j$ depends only on the capture status of the unit in
the previous $k$ occasions. 
More formally, for $k=1$ 
we have
that in order to uniquely specify the corresponding Markovian model 
we need to specify two probability parameters $\pi_{(0)} \in [0,1]$ and
$\pi_{(1)} \in [0,1]$  and set
\begin{eqnarray*}
& M_{c_{1}}: \begin{cases} 
p(X_{ij}=1 | x_{ij-1}=0)= \pi_{(0)}, \quad \forall i=1,\dots N \:\:
\forall j =1,\dots t \\
p(X_{ij}=1 | x_{ij-1}=1)= \pi_{(1)}, \quad \forall i=1,\dots N \:\: \forall j =2,\dots t  
\end{cases}	
\end{eqnarray*}
\normalsize 
while for $k=2$ we need to fix four parameters $\pi_{(00)}$, 
$\pi_{(01)}$, $\pi_{(10)}$ and $\pi_{(11)}$ and set 
\begin{eqnarray*}
&M_{c_{2}}:
\begin{cases}
Pr(X_{ij}=1 | x_{ij-2}=0,x_{ij-1}=0)= \pi_{(00)}, \quad \forall i=1,\dots N \:\: \forall j =1,\dots t\\
Pr(X_{ij}=1 | x_{ij-2}=0,x_{ij-1}=1)= \pi_{(01)}, \quad \forall i=1,\dots N \:\: \forall j =2,\dots t\\ 
Pr(X_{ij}=1 | x_{ij-2}=1,x_{ij-1}=0)= \pi_{(10)}, \quad \forall i=1,\dots N \:\: \forall j =3,\dots t\\
Pr(X_{ij}=1 | x_{ij-2}=1,x_{ij-1}=1)= \pi_{(11)}, \quad \forall i=1,\dots N \:\: \forall j =3,\dots t 
\end{cases}	
\end{eqnarray*}
For notational consistency we clarify that 
for $k=1,2$ if $j-k\leq 0$ the conditioning events related to
$x_{ij-k}$ are dropped. 
Indeed for the initial events in 
$M_{c_{1}} $ we conventionally assume that 
there has been no capture before the first occasion i.e. 
$p_1()\equiv Pr(x_{i1} = 1) = p_{2}(0) = \pi_{(0)}$
while 
in 
$M_{c_{2}} $ we conventionally assume that 
$p_1() \equiv Pr(x_{i1} = 1) = p_{3}(00) = \pi_{(00)}$ and also 
$p_2(0) \equiv Pr(x_{i2} = 1| x_{i1}=0) = p_{3}(00) = \pi_{(00)}$
and 
$p_1(1) \equiv Pr(x_{i2} = 1| x_{i1}=1) = p_{3}(01) = \pi_{(01)}$.
In the specific case
where $t=5$ the above Markovian models correspond to the following
partitions: for the first order we have
\begin{eqnarray}
\label{mc1}
\mathcal{H}_{2}(M_{c_{1}})=
\begin{cases}
H_1=\{ (),(0),(00),(10),(000),(100),(010),(110),  \\
 \qquad\quad(0000),(0100),(0010),(0110),(1000),(1100),(1010),(1110) \}  \\
H_2=\{ (1), (01),(11),(001),(101),(011),(111),  \\
\qquad\quad(0001),(0011),(0101),(0111),(1001),(1011),(1101)(1111) \}  
\end{cases}
\end{eqnarray}
and for the second order 
\begin{eqnarray}
\label{mc2}
\mathcal{H}_{4}(M_{c_{2}})=
\begin{cases}
H_1=\{ (),(0),(00),(000),(100),(0000),(0100),(1000),(1100) \} \\
H_2=\{ (10),(010),(110),(0010),(0110),(1010),(1110) \}  \\
H_3=\{ (1),(01),(001),(101),(0001),(0101),(1001),(1101) \}  \\
H_4=\{ (11),(011),(111),(0011),(0111),(1011),(1111) \}  
\end{cases}	
\end{eqnarray}
There are many other new models which can be derived from this
partition-based 
approach and some very simple examples are detailed  in the
Supplementary Web Materials (S1). 
Unfortunately, the number of all possible models is 
exponentially growing with $t$, 
namely with the Bell number of $2^{t}-1$, 
with more than $10^{25}$ alternatives when 
there are only $t=5$ capture occasions. 
Indeed, hardly all possible partitions lead to 
interesting behavioral  patterns or 
meaningful models.

In the following section we will introduce our main  idea of defining a
quantification of the partial capture history to be used as a 
covariate within a generalized linear model framework. Not only can
this covariate be used in its own right to define new parsimonious and
meaningful behavioural
models but it can also be used to partition the whole set of
conditional probabilities into equivalence classes recovering many
existing models as well as generating new meaningful ones.

\section{A new meaningful behavioural covariate approach}
\label{sec:covariate}
   
Similarly to \cite{Hugg:on:1989} and \cite{Alho:1990}, 
we consider a logistic regression model 
viewing each capture occurrence of unit $i$ at occasion $j$ as a binary outcome
whose probability can be modelled as a function of an explanatory
variable 
$z_{ij}$. 
Our idea is to build up and exploit a synthetic 
unidimensional 
$z_{ij}=q(x_{i1}, \dots ,x_{ij-1}) \in \Re$
associated to 
the previous partial capture history. 
Formally this can be embedded as follows
\begin{eqnarray}\mbox{logit} \left( p_{j}(x_{i1},\dots x_{ij-1}) 
\right) \equiv \log\left( 
\frac{Pr(X_{ij}=1|x_{i1},\dots x_{ij-1})}{1-Pr(X_{ij}=1|x_{i1},\dots x_{ij-1})}
\right)  = r(q(x_{1},\dots x_{j-1})) =  r(z_{ij})
\label{logistic}
\end{eqnarray} 
To begin with we consider a simple linear logistic regression for the probability of
each capture event $X_{ij}$, 
\begin{eqnarray}
 \mbox{logit} \left( 
p_{j}(x_{i1},\dots x_{ij-1}) 
\right) = \alpha + \beta z_{ij} \qquad \forall i=1,\dots N \:\: \forall j =1,\dots t
 \label{linear-logistic}
\end{eqnarray}
where $z_{ij}$ is a suitable numeric summary or quantification 
of a generic partial capture history $(x_{i1}\dots ,x_{ij-1})$.
To simplify the notation the unit index $i$ 
will be omitted in the following when it is not needed.
Remind that a partial capture history $\mathbf{x}=(x_{1},\dots ,x_{r})$
is a binary string taking values in
$H=\cup_{r=0}^{t-1}\mathcal{X}^{r}$ and 
has a length 
$l_{\mathbf{x}}=r$ which can take values in $0,1,...,t-1$.
Any partial capture history can be transformed  
into an integer number $z$ using the string 
as its binary representation and, after providing  some detail on 
this transformation, we will explain why this can be thought of as a
meaningful behavioural covariate. 
According to the natural and intuitive interpretation of grading a behavioural effect so that 
the occurrence of trapping in the last occasions 
has a greater impact on the future capture probability 
than those occurred in the previous ones we proceed to appropriately 
reverse the usual binary representation of integers and consider
the following transform
\begin{eqnarray*}
f(\mathbf{x}) = f(x_{1},\dots ,x_{l_{\mathbf{x}}}) = \sum_{j=1}^{l_{\mathbf{x}}} x_{j} 2^{j-1} \in \{0,1,2,...,2^{l_{\mathbf{x}}}-1\}
\end{eqnarray*}
where we assume that the partial capture history has length $l_{\mathbf{x}}\geq 1$. 
Conventionally, we set $f(\mathbf{x})=0$ for the empty binary sequence of length zero 
corresponding to $\mathbf{x}=()$.
However, note that the mapping $\mathbf{x} \mapsto f(\mathbf{x})$ 
spans a different range of integers 
$[0,2^{l_{\mathbf{x}}}-1]$ 
according to the length
$l_{\mathbf{x}}$ of the binary string $\mathbf{x}$. 
Hence, in order to obtain a potentially continuous covariate in a
fixed 
range to be used as 
a synthetic representation of the past history we rescale the range 
in the unit interval  by simply dividing $f(\mathbf{\mathbf{x}})$ by 
$2^{l_{\mathbf{x}}}-1$ and get our proposed numerical covariate $z$ 
\begin{eqnarray}
z &=& g(x_{1},\dots ,x_{l_{\mathbf{x}}}) = g(\mathbf{x}) = \frac{f(\mathbf{x})}{2^{l_{\mathbf{x}}}-1} \in \left\lbrace 0,\frac{1}{2^{l_{\mathbf{x}}}-1},\frac{2}{2^{l_{\mathbf{x}}}-1},...,1\right\rbrace 
\label{z}
\end{eqnarray}
From now on we will pretend that $z$ is a continuous time-varying covariate. 
As a matter of fact the function $g(\mathbf{x})$ has a finite-discrete range. 
However, if we extend $\mathbf{x}$ to be a possibly infinite sequence we have that
$\left\lbrace g(\mathbf{x}): \mathbf{x}\in H \right\rbrace $ corresponds to the set of dyadic rationals in $[0,1]$
which is a dense subset in $[0,1]$.\\
At first sight this may be thought of only as a technical mathematical device, 
but it can be easily argued that the ordering 
induced by this quantization of the previous binary history is sensible.
To begin with, 
the transformation $g(\mathbf{x})$ introduces a meaningful 
ordering of partial capture histories.
In fact one can argue that in the process of learning from the past experience a capture occurrence (1 digit) 
in the very last occasion (last position of the binary string) can affect the individual behaviour
with a greater impact than a capture in the previous occasions. 
Moreover, the more the capture occurrence (1 digits) in the partial capture history the greater the impact.
Of course we are not claiming the necessity of such ordering 
but we are explaining how it can be reasonably and fairly interpreted. 
Even though there is no compelling argument for the corresponding 
quantization it can be 
considered a convenient starting point to be refined further 
with alternative suitable
data-driven rescaling such as the one in \eqref{z} or other transformations.
In this sense
it can be given a plausibly realistic interpretation as 
a standardized quantization of the past experience or the accumulation of 
practice/memory/training 
with respect to the previously occurred events.
We will illustrate its usefulness to 
model behavioural effects in a capture-recapture context.

Considering 
 the partial capture histories corresponding 
to the capture occurrences in 
$\mathbf{X}=[x_{ij}]$ 
and 
the function $g:H\rightarrow [0,1]$ as in \eqref{z} 
one can derive a covariate matrix $\mathbf{Z}=[z_{ij}]$ as follows
\begin{eqnarray*}
z_{ij}=g(x_{i1},\dots ,x_{ij-1}) \quad \forall i=1,\dots ,N \: ; \: \forall j=1,\dots ,t.
\end{eqnarray*}
Notice that the first column of $\mathbf{Z}$ corresponds to a null column 
since, for $j=1$, the partial history $\mathbf{x}=(x_{i1},\dots x_{ij-1})$ 
corresponds in fact to an empty history ($\mathbf{x}=()$).\\
We now show in practice how the covariate mapping 
$\mathbf{x} \mapsto z$  works.
Consider the following complete capture history in a capture-recapture 
setting with $t=10$ trapping occasions:
$$
(x_{i1},\dots ,x_{i10}) \: = \: (0, 0, 1, 0, 0, 1, 1, 0, 0, 1) \\ 
$$
We derive all the quantizations corresponding to all
partial capture histories in Table~\ref{zzzz}
\begin{table}
\caption{\textit{Quantization of all partial capture histories corresponding
  to the complete individual capture history
  $(0,0,1,0,0,1,1,0,0,1)$ in a capture-recapture experiment
with $t=10$ capture occasions}} 
\label{zzzz}
\begin{center}
\begin{tabular}{|ccll|}
\hline \hline
\multicolumn{1}{|c}{\small{\textbf{Time}}} & \multicolumn{1}{c}{\small{\textbf{Current occurrence}}} & \multicolumn{1}{c}{\small{\: \textbf{Partial capture history}} } & \multicolumn{1}{c|}{\small{\textbf{Numeric covariate}} }\\
\multicolumn{1}{|c}{$j$} & \multicolumn{1}{c}{$x_{ij}$} & \multicolumn{1}{c}{$(x_{i1},\dots ,x_{ij-1})$} & \multicolumn{1}{c|}{$z_{ij}$} \\ 
\hline \hline 
\small{1} & \small{0}&  \small{$( \: \:) $}  & \small{ 0.000} \\ 
\small{2} & \small{0}&  \small{$( \: 0\: )$} & \small{ 0.000 = 0/1} \\ 
\small{3} & \small{1}&  \small{$( \: 0\:, \: 0\: )$} & \small{ 0.000 = 0/3}  \\ 
\small{4} & \small{0}&  \small{$( \: 0\:, \: 0\:, \: 1\: )$} & \small{ 0.571 = 4/7}  \\ 
\small{5} & \small{0}&  \small{$( \: 0\:, \: 0\:, \: 1\:, \: 0\: )$} & \small{ 0.267 = 4/15}  \\ 
\small{6} & \small{1}&  \small{$( \: 0\:, \: 0\:, \: 1\:, \: 0\:, \: 0\: )$} & \small{ 0.129 = 4/31}  \\ 
\small{7} & \small{1}&  \small{$( \: 0\:, \: 0\:, \: 1\:, \: 0\:, \: 0\:, \: 1\: )$} & \small{ 0.571 = 36/63}  \\
\small{8} & \small{0}&  \small{$( \: 0\:, \: 0\:, \: 1\:, \: 0\:, \: 0\:, \: 1\:, \: 1\: )$} & \small{ 0.787 = 100/127}  \\ 
\small{9} & \small{0}&  \small{$( \: 0\:, \: 0\:, \: 1\:, \: 0\:, \: 0\:, \: 1\:, \: 1\:, \: 0\: )$} & \small{ 0.392 = 100/255}  \\ 
\small{10} & \small{1}&  \small{$( \: 0\:, \: 0\:, \: 1\:, \: 0\:, \: 0\:, \: 1\:, \: 1\:, \: 0\:, \: 0\: )$} & \small{ 0.196 = 100/511} \\ 
\hline \hline
\end{tabular}
\end{center}
\end{table}
In our capture-recapture analysis we will use $z_{ij}$ 
as an individual covariate changing with time $j$.
For implementation purposes, both $\mathbf{X}$ and $\mathbf{Z}$ can be vectorized considering 
each double index $ij$ as a label for a single binary outcome $x_{ij}$ whose
probability can be explained in terms of the corresponding covariate $z_{ij}$.
In the following we will start considering a simple linear logistic model as in \eqref{linear-logistic}
but other more flexible models can be adopted such as
polynomial logistic regression, splines, step functions etc.
Notice that, differently from the usual covariates observable 
in a capture-recapture context 
during the experiment
(sex, age, length, etc.) 
we do know the values of the $z$'s also for the unobserved units. 
In fact, considering that
units observed are labelled from 1 to $M$ and 
those not observed are labelled from $M+1$ to $N$ we have
$z_{ij}=0$  for all  $i=M+1,\dots N$ and for all  $j=1,\dots ,t$.
We remark that other 
partial history 
orderings and 
mappings
can be considered sensible and useful 
in real data applications such as those based 
on the absolute or relative number of events experienced previously than 
time $j$. The reason why we are particularly interested in the ordering induced 
by $g(\mathbf{x})$ as in \eqref{z} is that it is a rather flexible 
device which can be also used to reproduce and flexibly modulate 
a Markov structure of arbitrary order. 
We will explain in detail the relationship between the continuous
covariate $z$ and the Markovian structure in subsection~\ref{z-markovian}.
We briefly illustrate alternative quantization of past experience in 
subsection~\ref{subsec:laternativecovariates}. \\
Finally notice that considering a numeric covariate $z$ built as described in \eqref{z} 
and a generic linear logistic regression model as in
\eqref{linear-logistic} 
the first capture probabilities turn out to be equal to 
$p_{1}()=p_{2}(0)=p_{3}(0,0)=\cdots =p_{t}(0,\dots ,0) = {e^\alpha}/{(1+e^\alpha)}$
and depend only on the parameter $\alpha$ while $\beta$ affects only  
the recapture probabilities which are indeed different according to
the different size of the memory effect as recorded by $z$.
This kind of model can then be considered an extension of the standard behavioral model $M_{b}$.
Moreover
the probability $P_{0}$ of never being captured during the whole experiment is
\begin{eqnarray*}
P_{0}=\left( 1-\frac{e^\alpha}{1+e^\alpha}\right)^t
\end{eqnarray*}
and depends only on one parameter as in model $M_b$.
Note however that, differently from $M_b$, 
also the recapture probabilities depend on $\alpha$ and this 
is the reason why the two models end up with different estimates of 
$P_0$ and $N$. 
In fact in
\cite{aft:2012} 
it is highlighted that 
all behavioural models for which 
the first equivalence class $H_1$ is composed exclusively 
by all partial capture
histories with no capture 
yield the same unconditional likelihood factor 
involving only $N$ and $p_{H_1}$ 
and they yield the same estimates for $N$.
For further details see 
Supplementary Web Materials (S1).

\subsection{Covariate representation and Markovian structure}
\label{z-markovian}
In this subsection we go back to the topic of building behavioural models
based on meaningful partitions of the subset $H$ as in
Section~\ref{sec:setup}.
We will show how the numeric covariate $z$ 
can be {\em also} used to set up meaningful partitions of $H$
and how one can recover those partitions corresponding to Markovian models.
If we fix a positive integer $k<t$ we can partition the set $H$ of all partial capture histories
according to the value of $g(\mathbf{x})$ into appropriate subintervals
namely 
$I_1 =\left[ 0,\frac{1}{2^k}\right]$, 
... ,
$I_{r} =\left( \frac{r-1}{2^k},\frac{r}{2^k}\right]$,
$I_{r+1} =\left( \frac{r}{2^k},\frac{r+1}{2^k}\right]$,
... ,
$I_{2^{k}} =\left( \frac{2^k-1}{2^k},1\right]$.
Hence we get the partition $H=\{H_{1},\dots ,H_{2^{k}}\}$ where
\begin{eqnarray}
\mathbf{x}\in H_{r} \: \Leftrightarrow \: z=g(\mathbf{x})\in I_{r} \quad \forall \: r \in \{1,\dots ,2^{k}\}
\label{***}
\end{eqnarray}
so that the equivalence classes of binary subsequences
depend only on the last $k$ binary events.
In fact one can formally show that the mapping $g$ defined in \eqref{z} is such that,
for each partial capture history $\mathbf{x} \in H$,
$z=g(\mathbf{x})$ belongs to the same set $I_r$
according to the last $k$ digits of the binary sequence.
Hence the definition of 
equivalence classes of 
conditional probabilities
given partial capture histories in these partitions
satisfy the Markov property of order $k$.
The formal proof is provided in 
Supplementary Web Materials (S2)
with further 
details on ensuring the appropriate correspondence 
of partitions $H_1,...,H_{2^k}$ with 
$I_1$,...,$I_{2^{k}}$
deriving from \eqref{***}
also for partial capture histories with less than $k$ digits.

We highlight that the partition defined in \eqref{***} is equivalent to considering a general logistic regression as in \eqref{logistic}
where the function $r(z)$ is a real step-function  $s(z)$ which is constant 
over each subinterval $I_{r}$ 
as follows
\begin{eqnarray*}
s(z)=logit(p_{H_{r}}) \qquad z \in I_{r}
\qquad  \forall r=1,...,2^k
\end{eqnarray*}
where $p_{H_{r}}=P(X_{ij}=1 | x_{i1},\dots ,x_{ij-1}))$ 
for any $(x_{i1},\dots ,x_{ij-1})\in H_{r}$
according to the notation used in 
Section~\ref{sec:setup}.
Notice that 
in the logistic regression setup 
this is equivalent 
to convert the 
numerical covariate into a 
categorical factor 
according to which subinterval $I_r$ the 
covariate falls in. 

In order to get it straight  we illustrate the first order Markov case 
with $k=1$ and the corresponding two subintervals 
$I_1=\left[0,\frac{1}{2}\right]$, 
$I_2=\left(\frac{1}{2},1\right]$
which divide the 
unit interval representing the support of the variable $z$.
From \eqref{***}  we get 
the partition of the set $H$ of all partial capture histories 
considered in the particular case with $t=5$
as in Table~\ref{tab:x}.  
The bipartition obtained is exactly the same as \eqref{mc1}
introduced in the previous section. 
More details and examples of the correspondence are 
included in the 
Supplementary Web Materials (S2).

\begin{table}[ht]
\caption{\textit{Capture-recapture experiment with $t=5$ capture
    occasions: list of all the possible partial capture histories ($\mathbf{x}$)
    and their corresponding numerical covariates ($g(\mathbf{x})$)
    with relative subintervals for 
   Markov models of order $k=1$ and $k=2$} }
\label{tab:x}
\begin{center}{\footnotesize
\begin{tabular}{|lllllllll|}
\hline \hline
                      &                            &\multicolumn{1}{c}{{\bf Partition}}
                      & \multicolumn{1}{c}{{\bf Partition}} & &
                      &                            & \multicolumn{1}{c}{{\bf Partition}} & \multicolumn{1}{c|}{{\bf Partition}}\\
$\mathbf{x}$ & $g(\mathbf{x})$  &\multicolumn{1}{c}{{\bf interval}}
&\multicolumn{1}{c}{{\bf interval}}  &  \phantom{mmmm} &
$\mathbf{x}$ & $g(\mathbf{x})$  &\multicolumn{1}{c}{{\bf interval}}   &\multicolumn{1}{c|}{{\bf interval}} \\
                      &                            & \multicolumn{1}{c}{($k=1$)}   &  \multicolumn{1}{c}{($k=2$)} & &
                      &                            &  \multicolumn{1}{c}{($k=1$)} & \multicolumn{1}{c|}{($k=2$)}\\
\hline \hline
() & 0 & $[0,0.5]$                & $[0,0.25]$    & &
(0) & 0 & $[0,0.5]$              & $[0,0.25]$     \\
(1) & 1 & $(0.5,1]$              & $(0.75,1]$    & &
(00) & 0 & $[0,0.5]$            & $[0,0.25]$    \\
(10) & 0.333 & $[0,0.5]$     & $(0.25,0.5]$    & &
(01) & 0.667 & $(0.5,1]$     & $(0.5,0.75]$    \\
(11) & 1 & $(0.5,1]$            & $(0.75,1]$    & &
(000) & 0 & $[0,0.5]$          & $[0,0.25]$    \\
(100) & 0.143 & $[0,0.5]$   & $[0,0.25]$    & &
(010) & 0.286 & $[0,0.5]$   & $(0.25,0.5]$    \\
(110) & 0.429 & $[0,0.5]$   & $(0.25,0.5]$    & &
(001) & 0.571 & $(0.5,1]$   & $(0.5,0.75]$    \\
(101) & 0.714 & $(0.5,1]$   & $(0.5,0.75]$    & &
(011) & 0.857 & $(0.5,1]$   & $(0.75,1]$       \\
(111) & 1 & $(0.5,1]$          & $(0.75,1]$    & &
(0000) & 0 & $[0,0.5]$        & $[0,0.25]$    \\
(1000) & 0.067 & $[0,0.5]$ & $[0,0.25]$     & &
(0100) & 0.133 & $[0,0.5]$ & $[0,0.25]$    \\
(1100) & 0.200 & $[0,0.5]$ & $[0,0.25]$    & &
(0010) & 0.267 & $[0,0.5]$ & $(0.25,0.5]$ \\
(1010) & 0.333 & $[0,0.5]$ & $(0.25,0.5]$ & &
(0110) & 0.400 & $[0,0.5]$ & $(0.25,0.5]$  \\
(1110) & 0.467 & $[0,0.5]$ & $(0.25,0.5]$ & &
(0001) & 0.533 & $(0.5,1]$ & $(0.5,0.75]$  \\
(1001) & 0.600 & $(0.5,1]$ & $(0.5,0.75]$ & &
(0101) & 0.667 & $(0.5,1]$ & $(0.5,0.75]$  \\
(1101) & 0.733 & $(0.5,1]$ & $(0.5,0.75]$ & &
(0011) & 0.800 & $(0.5,1]$ & $(0.75,1]$    \\
(1011) & 0.867 & $(0.5,1]$ & $(0.75,1]$    & &
(0111) & 0.933 & $(0.5,1]$ & $(0.75,1]$     \\
(1111) & 1 & $(0.5,1]$        & $(0.75,1]$    & &       
&&& \\
\hline \hline
\end{tabular}} 
\end{center}
\end{table}

\subsection{Alternative covariate partitioning and alternative meaningful behavioural covariates} 
\label{subsec:laternativecovariates}

We now sketch a list of other meaningful alternatives for partitioning
the {\em covariate range}. 
Indeed, it is possible to recover model $M_{b}$ 
associated to the partition $\mathcal{H}_{2}(M_{b})$ 
by partitioning the support of $z=g(\mathbf{x})$ as follows
$I_1=\left[ 0,\frac{1}{2^{t}}\right]$, 
$I_2=\left(\frac{1}{2^{t}},1\right]$
so that it can be recovered in terms of 
the logistic regression 
with step function defined as 
$s(z)=logit(p_{H_{1}})  = logit(p)$ when  $z \in I_1$
and 
$s(z)=logit(p_{H_{2}})  = logit(r)$ when  $z \in I_2$. 
The upper bound of 
$I_1=\left[ 0,\frac{1}{2^{t}}\right]$ is chosen conveniently low 
in order to get the same partition $\mathcal{H}_{2}(M_{b})$.
In fact, 
the presence of at least one capture in a partial capture
history $\mathbf{x}$ 
makes the corresponding 
$g(\mathbf{x}) \geq 1/2^{t-1} > 1/2^t$.
Notice that 
since $ 1/2^t>0$ 
the first partition $I_1$ can be equivalently reduced to
the single value $\{0\}$ or any other interval 
$[ 0,e_1]$ provided that $e_1 \leq 1/2^t$. This is basically 
due to the discreteness of the observable range. 

More generally,  
an alternative partition 
of the range of $z$ into $A$ consecutive 
subintervals $I_{1} = [0,e_{1}]$,  
$\dots$, $I_{a} = (e_{a-1},e_a]$, $\dots$, 
$I_{A}=(e_{A-1},1]$
represents a meaningful 
behavioural model corresponding to the 
regression step function 
\begin{eqnarray*}
s(z)=logit(p_{H_{a}}) \qquad \forall z \in I_a  \qquad 
 a=1, \dots A.
\end{eqnarray*}

This particularly 
flexible instance of partitioning the range 
of the behavioural covariate $g(\mathbf{x})$
embeds some of the original models  
proposed in \cite{Farc:2011} such as $M_{L_2}$ 
(Supplementary Web
Materials, S1 and S3).
In fact looking for an appropriate number and location of the
partition 
cuts $e_a$ 
in terms of step functions for the logistic regression 
can readily explore a range of
meaningful variable order Markov chain
models. We will exploit this approach in our applications.

As already mentioned the most critical parameter for the estimation of
$N$ is the probability 
$P_{0}$ as in \eqref{P0}. 
Indeed when we partition the set of 
conditional probabilities 
through partitioning  
the quantification 
$z=g(\mathbf{x})$ 
into intervals 
$I_1$, ... , $I_{A}$ 
we have that, as long as 
$ g(\mathbf{x}) \in I_1 =[0,e_1]$ (and this is certainly true   
for all partial capture histories with no capture)
we get 
$$
p_{j}(0,\dots ,0) = p_{H1} \qquad \forall j=1,...,t-1
$$
so that the fundamental probability 
$P_{0}=\left[(1-p_{1}())\prod_{j=2}^{t}(1-p_{j}(0,\dots ,0)) \right] = 
\left(1-p_{H_1} \right)^t$ depends on a single element 
(first component) of the parameter vector 
$\mathbf{p}_{H_A}$.


As previously highlighted, the procedure of ordering and scaling a generic 
partial capture history defined in \eqref{z}
is not the only way of representing the quantization of a binary sequence.
Indeed, although we have argued 
why our
choice of $g(\mathbf{x})$ can be considered reasonable 
in some cases (also in terms of Markovian structure) it can be open to some criticism.
For example, consider the following two partial capture histories each
based on five capture occasions 
\begin{align*}
&\mathbf{x}_{1}=(1,1,1,1,0) \quad \Rightarrow \quad g(\mathbf{x}_{1})=\frac{1\cdot 2^0+1\cdot 2^1+1\cdot 2^2+1\cdot 2^3+0\cdot 2^4}{2^5 -1}=\frac{15}{31}=0.484\\
&\mathbf{x}_{2}=(0,0,0,0,1) \quad \Rightarrow \quad g(\mathbf{x}_{2})
= \frac{0\cdot 2^0+0\cdot 2^1+0\cdot 2^2+0\cdot 2^3+1\cdot 2^4}{2^5 -1}=\frac{16}{31}=0.516
\end{align*}
The first partial capture history $\mathbf{x}_{1}$ 
has a total  of four captures 
in the first four occasions 
while the second one $\mathbf{x}_{2}$ has only one capture in the last occasion.
The mapping $g(\cdot)$ described in \eqref{z} assigns a larger impact on the conditional probabilities 
to $\mathbf{x}_{2}$.
One can find undesirable the fact that 
the partial capture history 
$\mathbf{x}_{1}$
having just a single capture, 
even though in the last occasion, yields a larger value compared to 
a binary sequence which has 4 captures out of 5.\\
As a possible alternative useful mapping one can consider a function based on 
the {\em total}  number of captures 
occurred for each partial capture history $\mathbf{x} \in H$.
In order to obtain a potentially continuous covariate as in \eqref{z} 
we rescale the range in the unit interval considering as denominator
the length of each capture history as follows
\begin{align}
z=g_{n}(\mathbf{x})=g_{n}(x_{1},\dots ,x_{l_{\mathbf{x}}})=\frac{\sum_{j=1}^{l_{\mathbf{x}}}x_{j}}{l_{\mathbf{x}}} 
\in \left\lbrace 0,\frac{1}{l_{\mathbf{x}}},\frac{2}{l_{\mathbf{x}}},...,1\right\rbrace 
\label{z_1}
\end{align}
The partial capture histories $\mathbf{x}_{1}$ and $\mathbf{x}_{2}$
can be quantified as in \eqref{z_1} yielding 
$\mathbf{x}_{1}=(1,1,1,1,0)\mapsto  \frac{4}{5}=0.8$ and 
$\mathbf{x}_{2}=(0,0,0,0,1) \mapsto \frac{1}{5}=0.2$.
It is also possible to rescale the number of captures 
by considering the total number of occasions in the whole experiment as follows
\begin{align}
z=\tilde{g}_{n}(\mathbf{x})=\tilde{g}_{n}(x_{1},\dots ,x_{l_{\mathbf{x}}})=\frac{\sum_{j=1}^{l_{\mathbf{x}}}x_{j}}{t} 
\in \left\lbrace 0,\frac{1}{t-1},\frac{2}{t-1},...,1\right\rbrace 
\label{z_2}
\end{align}
On the other hand the mapping $g_{n}$ and $\tilde{g}_{n}$ described in \eqref{z_1} and \eqref{z_2} 
may have in turn their own undesirable features.
In fact they do not take into account the inner sequence structure 
considering the number of captures only. 
For example a partial capture history (1,0,0,0,0) will be equivalent
in terms of $g_{n}$ and $\tilde{g}_{n}$ to $\mathbf{x}_{2}$ even though they may be considered substantially different.

\section{Unconditional maximum likelihood inference}
\label{sec:inference}

In this section we show how our 
new approach exploiting a numerical summary of partial 
capture histories and a logistic regression framework
yields a simple-to-implement procedure to infer on 
the parameter space through the unconditional likelihood
and,  as a by-product,  inference on the main parameter of interest $N$ 
using the profile likelihood. 
Indeed we will basically recycle 
consolidated standard GLM routines in our capture-recapture context. 
Let $L(N,\alpha ,\beta)$ be the likelihood function for 
the linear logistic model (\ref{logistic}) such that
\begin{eqnarray}
L(N,\alpha ,\beta)\propto \binom{N}{M} \left[ \prod_{i=1}^{N}\prod_{j=1}^{t} \left( \frac{\exp(\alpha + \beta z_{ij})}{1+\exp(\alpha + \beta z_{ij})}\right)^{x_{ij}} \left( 1- \frac{\exp(\alpha + \beta z_{ij})}{1+\exp(\alpha + \beta z_{ij})}\right)^{1-x_{ij}}\right]  
\label{lik.mbc}
\end{eqnarray}
In order to make inference on $N$
one can first look at $L(N,\alpha ,\beta)$ as a function of
$(\alpha,\beta)$ only for a fixed value of $N$.
Let us denote with 
$$
\hat{L}(N)=L(N,\hat{\alpha}(N),\hat{\beta}(N))=\sup_{\alpha ,\beta}L(N, \alpha ,\beta)
$$
the maximum likelihood 
of $(\alpha,\beta)$
obtained as a result of 
a standard logistic model fitted using 
$N\times t$ binary observations $x_{ij}$ 
with their corresponding numerical covariates $z_{ij}$. 
Unconditional maximum likelihood estimate for $N$ will then be
\begin{eqnarray*}
\hat{N}=\arg\max_{N\in  \{M,\dots ,N_{upp}\}} \left( \hat{L}(N)\right) 
\end{eqnarray*}  
where $N_{upp}$ is a suitably high fixed upperbound for the population
size.
The joint unconditional likelihood for all parameters 
involved in the model is globally maximized at 
the UMLE value
$(\hat{N},\hat{\alpha}(\hat{N}),\hat{\beta}(\hat{N}))$.
Hence  the estimating procedure for obtaining 
maximum of the unconditional likelihood function 
$L(N,\alpha ,\beta)$
requires to iteratively fit a logistic regression
for each $N\in  \{M,\dots ,N_{upp}\}$. 
For very large values of $N_{upp}$ this procedure can 
be computationally demanding
and time-consuming 
involving logistic procedures repeated $N_{upp}-M+1$ times.
To reduce computational effort and computing time 
it is possible to 
group observed results 
according to the same value of the covariate
such as those corresponding to unobserved units and 
implement GLM routines for weighted data. 
Moreover, one can 
evaluate the profile likelihood function 
not at each single value of $N \in \{M,\dots ,N_{upp}\}$ 
but only on a suitable sub grid 
and use some parallel computing environment to run 
simultaneously multiple logistic fits. 
Standard GLM routines also allow to fit 
more flexible models incorporating unobserved 
heterogeneity adding on a logit scale 
an individual random effect to the probability of the 
longitudinal series of $t$ binary outcomes.



\section{Examples}
\label{sec:examples}

{\em Great Copper Butterflies}. 
As a first example we will 
consider the Great Copper data originally analyzed in \cite{rams:seve:2010}
and also reviewed in \cite{Farc:2011} and \cite{aft:2012}. 
There are $t=8$ capture occasions and $M=45$ observed butterflies. 
\cite{rams:seve:2010} 
explain that butterflies tend to congregate near favorable habitat, which is readily recognized
by the observer and this may yield a persistence related to the characteristics
of the subject animals, the environment and/or the observational
pattern. In the first attempt 
to model and understand the dependence
structure in the data \cite{rams:seve:2010} show how there can be a great impact of
the modelled temporal dependence and behavioural pattern on
the final estimates with possible large uncertainty on the magnitude
of the population size. In fact \cite{Farc:2011} and \cite{aft:2012}
provided evidence of alternative ephemeral effects which 
correspond to possibly larger population estimates. We now show the
ability of the behavioral covariate approach to improve model fit
and gain an alternative 
simple parsimonious understanding of the dependence pattern.
From the maximization of the unconditional likelihood 
we get the results displayed in Table~\ref{tab:2} where models are
ranked according to the increasing value of the AIC. We have
considered as competing models 
a linear logistic model
as in \eqref{linear-logistic} with $z=g(\mathbf{x})$ denoted with
$M_{z}$ and,  with similar notation, linear logistic models
$M_{z_{g_n}}$, $M_{z_{f}}$ and  $M_{z_{\tilde{g}_n}}$ 
with, respectively,  $z=g_n(\mathbf{x})$, $z=f(\mathbf{x})$ 
and $z=\tilde{g}_{n}(\mathbf{x})$.
Moreover, model $M_{c_kb}$ is a 
$k$-th order Markovian model with a specific first capture probability 
which differs from the recapture probability conditioned on the
absence of capture in the last $k$ occasions
(see also Supplementary Web Materials (S1)). 

From results displayed in 
Table~\ref{tab:2}
it is apparent that
the use of the behavioural covariate $z=g(\mathbf{x})$ allows for a 
sensible improvement of the AIC which is however accompanied by a 
larger point estimate and width of the confidence interval.
More precisely,
model $M_{z}$  
yields 
$\hat{N}=170$, 
with $\hat{\alpha}= -3.243$
$\hat{\beta}=3.179$.
A significantly positive $\hat{\beta}$ ($p$-value $<10^{-10}$) 
highlights an initial trap-happiness effect 
which tends to diminish  when the memory effect covariate 
$z=g(\mathbf{x})$ decreases. 
 
We have also implemented the idea of partitioning 
the range of the meaningful covariate in order to look for further
improvements. 
In fact one can see from Table~\ref{tab:2}  that, 
although we were not able to 
improve the AIC of $M_z$, 
the best model 
we could  fit 
looking for an 
appropriate number of optimal cutpoints  
(ranging from 1 to 4) 
is $M_{z.cut(4)}$. 
Model $M_{z.cut(4)}$ 
corresponds to 
the partition of the covariate $z=g(\mathbf{x})$ with 
$I_1=[0,1/2^8]$, 
$I_2=(1/2^8,0.250]$,
$I_3=(0.254,0.571]$, 
$I_4=(0.571,0.857]$, 
$I_5=(0.857,1]$.
Notice that the first interval $I_1$
determines the same first partition subset $H_1$ of the partition 
$\mathcal{H}_{2}(M_{b})$
in 
\eqref{Mb-part}
corresponding to the classical 
behavioural 
model.
In this case, as already argued in 
\cite{aft:2012},
the unconditional MLE 
yields the same estimate 
for $N$ in both models 
although with a very different 
AIC index.
For the  conditional probability parameter
estimates
we get 
$\hat{p}_{H_1}=0.147$,
$\hat{p}_{H_2}= 0.024 $,
$\hat{p}_{H_3}=0.134 $,
$\hat{p}_{H_4}=0.273 $ and 
$\hat{p}_{H_5}=0.500$.
This pattern as well as that resulting from 
$M_{z}$ 
could be interpreted as an 
initial trap-happiness response 
(from $\hat{p}_{H_1}$ to
$\hat{p}_{H_5}$) 
followed by 
decreasing recapture probabilities
$(\hat{p}_{H_5}<\hat{p}_{H_4}<\hat{p}_{H_3}<\hat{p}_{H_2} 
<\hat{p}_{H_1})$
vanishing with the decreasing memory effect corresponding to the 
covariate $z$. 

These results show that, although 
the enduring effect of 
the classical behavioural model 
yields one of the 
worst fitting models, 
a novel 
mixed  
ephemeral-enduring 
behavioral effect 
is highlighted 
by modelling 
the subsequent changes 
in the 
longitudinal pattern
of capture probabilities
after the first capture
by means of our 
meaningful behavioural covariate 
$z=g(\mathbf{x})$.
This model could
confirm a kind of persistence effect conjectured by 
\cite{rams:seve:2010}
although
there may remain some doubts on the ability of detecting the right 
longitudinal pattern with so few captured individuals during a
moderate number of trapping occasions. This issue will
be addressed in Section~\ref{sec:simulation}. 

\begin{table}
\caption{\textit{Great Copper Butterfly data: point and interval
    estimates together with AIC index of alternative fitted 
    models.  
Confidence intervals at level $1-\alpha=0.95$. 
We note that 
model $M_{z.cut(1)}$ 
corresponding to the
optimal single cut 
$e_1^*=0.625$
corresponds to  model 
$M_{L_2}$ 
described in
\citet{Farc:2011}}}
\label{tab:2}
\begin{center}
\begin{tabular}{|l|c|r|c|r|}
\hline \hline
\textbf{Model}  &  \# \textbf{parameters}   &    $\hat{N}$    &
$(N^{-},N^{+})$   &    \textbf{AIC}        \\
\hline \hline
$M_{z}$         &  1+2   & 170 & (87,448)  & 321.46 \\ 
$M_{z.cut(4)}$   &  1+5  & 62    & (48,223) & 321.54 \\ 
$M_{z.cut(3)}$   &  1+4  & 62    & (48,223) & 321.62 \\ 
$M_{z.cut(2)}$   &  1+3  & 176  & (78,243) & 323.36 \\ 
$M_{c_{2}b}$    & 1+5  & 62    & (48,223) & 325.46 \\ 
$M_{z_{g_{n}}}$  &  1+2   & 154 & (82,367)  & 325.99 \\ 
$M_{z.cut(1)}$    & 1+2 & 90     & (63,152) & 326.01 \\ 
$M_{c_{2}}$     & 1+4 & 176   & (78,896) & 327.20 \\ 
$M_{c_{1}}$     & 1+2 & 97     & (64,181) & 330.93 \\ 
$M_{c_{1}b}$   & 1+3  & 62    & (48,223) & 331.24 \\ 
$M_{z_{\tilde{g}_{n}}}$ &  1+2   & 96 & (62,184)  & 338.30 \\ 
$M_{0}$         & 1+1 & 64     & (53,85) & 342.80 \\ 
$M_{z_{f}}$  &  1+2   & 68 & (54,97)  & 343.77 \\ 
$M_{b}$         & 1+2 & 62     & (48,223) & 344.77 \\ 
$M_{t}$          & 1+8 & 64     & (52,84) & 352.85 \\ 
\hline \hline
\end{tabular} 
\end{center}
\end{table}

\medskip

{\em Giant Day Geckos}.  The giant day gecko 
(Phelsuma madagascariensis grandis) is a tropical
reptile 
living in areas of tropical and subtropical forest in northern Madagascar.
A capture-recapture sampling 
on the giant day gecko has been conducted
in the Masoala rainforest exhibit at the Zurich Zoo
and the resulting data have been analyzed in \cite{wang:etal:2009}.
Due to the high number of capture occasions ($t=30$) it can be considered
an unusually good dataset where the closed population assumption is valid
since it is a captive population.
We are interested in analyzing behavioural
patterns possibly originated  by 
feed habits of the geckos and/or by 
the human presence. More details on the 
sampling process
can be found in \cite{wang:etal:2009}.
In Table~\ref{tab:3} we list the results obtained by 
fitting a collection of standard  models as well as new models based on 
both the originally proposed $g(\mathbf{x})$ and $g_n(\mathbf{x})$.
With this data set we found that the alternative behavioural covariate
$z=g_n(\mathbf{x})$ based on the number of 
previous captures as defined in \eqref{z_1}
allows to achieve a better fit. 
In this example the AIC index highlights a 
different kind of behavioural response selecting as best model 
$M_{z_{g_n}.cut(3)}$ which partitions with three cutpoints the behavioural covariate
range in four subintervals 
$[0,0.05]$,
$(0.05,0.1579]$,
$(0.1579,0.625]$ 
and 
$(0.625,1]$
with corresponding conditional probability
estimates 
$\hat{p}_{H_1}=0.034$,
$\hat{p}_{H_2}= 0.061 $,
$\hat{p}_{H_3}= 0.159 $ and 
$\hat{p}_{H_4}=  0.375$.
The optimal cuts have been determined by a grid search as 
detailed in the  Supplementary Web Materials (S3).

\begin{table}[ht]
\caption{\textit{Giant Day Gecko data: point and interval
    estimates together with AIC index of alternative fitted
    models.  Confidence intervals at level $1-\alpha=0.95$. Linear logistic models 
as in \eqref{linear-logistic} with $z=g(\mathbf{x})$ is 
denoted with $M_{z}$; with $z=g_n(\mathbf{x})$ is 
denoted with $M_{z_{g_n}}$. Model $M_{c_kb}$ are
$k$-th order Markovian models with a specific first capture probability 
which differs from the re-capture probability conditioned on the
absence of capture in the last $k$ occasions}}
\label{tab:3}
\begin{center}
\begin{tabular}{|l|c|r|c|r|}
\hline \hline
\textbf{Model}  &  \# \textbf{parameters}   &    $\hat{N}$    &
$(N^{-},N^{+})$ & \textbf{AIC}   \\
\hline \hline
$M_{z_{g_n}.cut(3)}$ & 4+1 & 105 & (83,154) & 1108.76 \\ 
  $M_{z_{g_n}.cut(2)}$ & 3+1 & 89 & (77,108) & 1110.80 \\ 
  $M_{z_{g_n}.cut(1)}$ & 2+1 & 89 & (77,108) & 1114.81 \\ 
  $M_{z_{g_n}}$ & 2+1 & 86 & (76,101) & 1126.36 \\ 
  $M_{z_{\tilde{g}_n}}$ & 2+1 & 87 & (76,105) & 1141.09 \\ 
  $M_z$ & 2+1 & 80 & (73,91) & 1147.36 \\ 
  $M_{c_{2}b}$ & 5+2 & 107 & (79,266) & 1150.25 \\ 
  $M_{c_{1}b}$ & 3+1 & 107 & (79,266) & 1153.18 \\ 
  $M_{c_{2}}$ & 4+1 & 79 & (72,89) & 1154.70 \\ 
  $M_{b}$ & 2+1 & 107 & (79,266) & 1155.73 \\ 
  $M_{c_{1}}$ & 2+1 & 76 & (71,85) & 1160.32 \\ 
  $M_{t}$ & 30+1 & 74 & (70,81) & 1164.72 \\ 
  $M_{0}$ & 1+1 & 74 & (70,82) & 1166.18 \\ 
  $M_{z_{f}}$ & 2+1 & 75 & (70,82) & 1166.88 \\   
\hline \hline
\end{tabular}
\end{center}
\end{table}

\section{A simulation study for model selection}
\label{sec:simulation}

In this section a simulation study 
is developed in order to evaluate and compare the performance 
among alternative classical ($M_0$, $M_b$, $M_t$, $M_{c_1}$,
$M_{c_1b}$, $M_{c_2}$ 
and $M_{c_2b}$) 
and new ($M_{z}$, $M_{z_{g_n}}$, $M_{z_f}$ 
and $M_{z_{\tilde{g}_n}}$) models based on four meaningful behavioural
covariates.

Motivated by the real data results 
we decided to focus on models $M_z$ and $M_{z_{g_n}}$
which represent
different aspects of the behavioural effect to capture.
We use either one 
as generating data model with 
the same parameter settings ($\alpha = -3$ and $\beta = 4$)
considering different values for the population size and the number of occasions:
$N\in\{100,200\}$ and $t\in\{10,20,30\}$ respectively. 
Notice that, 
taking the same value of $t$ 
the probability $P_0$ of never being observed 
will be the same in  both models
and taking the same $N$ we get the same expected 
number $E[M]$ of distinct units captured at least once.
Obviously if $N$ and/or $t$ increase the expected number 
of distinct units observed becomes larger.

For each setting described in Table~\ref{tab:4} 
$K=100$ data-set are generated and for each generated data set we calculate point and interval estimates
using the unconditional likelihood approach. 
Moreover, the AIC index is computed in order to compare all candidate models.
In Tables~\ref{tab:5a} and \ref{tab:5b}, for all the alternative models considered, 
we report the empirical mean and the root mean square error (RMSE)
of the alternative estimates of $N$, 
empirical coverage and average length 
of the interval estimates and the percentage of times 
that each model is selected as the the best one when the AIC index
is used.
As we can see from the results in  Tables~\ref{tab:5a} and \ref{tab:5b}
the estimates of $N$ from the true model ($M_z$ and $M_{z_{g_n}}$ respectively)
almost always yield
best results in terms of both point and interval estimates.
In correspondence of the true model 
the empirical mean $\bar{N}$ is very close to the real values of $N$
and the RMSE is almost always the smallest one.
Only in Trial 1 and Trial 7 the true model does not achieve 
the smallest RMSE. However
they are the only ones which 
yield confidence intervals  
guaranteeing a coverage
close to the nominal level. 
Finally, 
from the column labelled with \%aic
one can verify 
that
the AIC index allows to identify 
the true model most of times, especially
when $t$ increases.
In fact, with a long sequence 
the longitudinal information gathered from the experiment is high and
the selection criterion is able to well distinguish among all candidates.
This is still true  also when alternative 
behavioural effects can be somehow related
as in the case of higher order Markovian models and 
model $M_z$.
However, when the number of capture occasions is low
and the number of distinct units is not too high  
the available information
could be not sufficient to correctly select the true model.  


\begin{table}[ht]
\caption{\textit{Description of simulation settings: in both
    generating models $M_{z}$  and $M_{z_{g_{n}}}$ logistic regression
    parameter values were set equal to $\alpha=-3$, $\beta=4$. The
    expected value of distinct observed units is denoted with $E[M]$ 
and can be computed as $E[M]=N(1-P_{0})$}}
\label{tab:4}
\begin{center}
\begin{tabular}{|c|l|r|r|r|}
\hline \hline
\textbf{Trial} & \textbf{Model} & $N$ & $t$ & $E[M]$ \\ 
\hline \hline
1& $M_{z}$ & 100 & 10 & 38.5 \\ 
 2& $M_{z}$ & 100 & 20 & 62.2 \\ 
 3& $M_{z}$ & 100 & 30 & 76.7 \\ 
 4& $M_{z}$ & 200 & 10 & 77.0  \\ 
 5& $M_{z}$ & 200 & 20 & 124.3 \\ 
 6& $M_{z}$ & 200 & 30 & 153.4 \\ 
 7& $M_{z_{g_{n}}}$ & 100 & 10 & 38.5  \\ 
 8& $M_{z_{g_{n}}}$ & 100 & 20 & 62.2 \\ 
 9& $M_{z_{g_{n}}}$ & 100 & 30 & 76.7 \\ 
 10&$M_{z_{g_{n}}}$  & 200 & 10 & 77.0  \\ 
 11&$M_{z_{g_{n}}}$  & 200 & 20 & 124.3 \\ 
 12&$M_{z_{g_{n}}}$  & 200 & 30 & 153.4 \\ 
\hline \hline
\end{tabular} 
\end{center}
\end{table}

\begin{sidewaystable}
\vspace*{15cm}\centering{
\begin{tabular}{|lccccccccccccccccc|}
\hline
 &
    \multicolumn{5}{c}{{\textbf{Trial} 1}}&&
    \multicolumn{5}{c}{{\textbf{Trial} 2}}&&
    \multicolumn{5}{c|}{{\textbf{Trial} 3}}\\
    \hline
{\textbf{Model}} 
& {$\bar{N}$} & {\textbf{rmse}} & {\textbf{CI}} & {$l_{CI}$} & {\textbf{\%aic}\:\:}  &
& {$\bar{N}$} & {\textbf{rmse}} & {\textbf{CI}} & {$l_{CI}$} & {\textbf{\%aic}\:\:} &
& {$\bar{N}$} & {\textbf{rmse}} & {\textbf{CI}} & {$l_{CI}$} & {\textbf{\%aic}} \\
\hline
  $M_{z}$ & 109 & 43.0 & 90 & 187.1 & 75 &
  & 100 & 13.9 & 95 & 52.6 & 88 &
  & 99 & 7.6 & 95 & 29.2 & 95\\
   $M_{z_{g_n}}$ & 75 & 33.9 & 66 & 78.8 & 2 &
   & 82 & 20.8 & 47 & 28.8 & 0 &
   & 87 & 13.9 & 34 & 16.4 & 0\\
  $M_{z_f}$ & 47 & 53.3 & 1 & 17.1 & 0 &
  & 67 & 33.4 & 0 & 8.7 & 0 &
  & 79 & 21.7 & 0 & 4.6 & 0\\
  $M_{z_{\tilde{g}_n}}$ & 57 & 45.2 & 21 & 40.3 & 0 &
  & 74 & 27.3 & 10 & 19.1 & 0 &
  & 83 & 17.8 & 7 & 11.4 & 0\\
  $M_{0}$ & 44 & 56.1 & 0 & 11.1 & 0 &
  & 66 & 34.3 & 0 & 7.1 & 0 &
  & 78 & 22.1 & 0 & 3.9 & 0\\
  $M_{b}$ & 209* & 564.0* & 83 & 3866.0* & 0 &
   & 119* & 105.4* & 89 & 1776.5* & 0 &
   & 98 & 15.3 & 90 & 79.3 & 0\\
   $M_{c_1}$ & 72 & 34.7 & 62 & 82.1 & 3 &
   & 83 & 19.6 & 55 & 30.2 & 0 &
   & 89 & 12.4 & 50 & 18.4 & 0\\
   $M_{c_1b}$ & 209* & 564.0* & 83 & 3866.0* & 0 &
   & 119* & 105.4* & 89 & 1776.5* & 0 &
   & 98 & 15.3 & 90 & 79.3 & 0\\
   $M_{c_2}$ & 96* & 46.2* & 89 & 151.7* & 14 &
   & 93 & 15.9 & 86 & 47.0 & 9 &
   & 94 & 9.1 & 84 & 25.3 & 4\\
   $M_{c_2b}$ & 209* & 564.0* & 83 & 3866.0* & 6 &
   & 119* & 105.4* & 89 & 1776.5* & 3 &
   & 98 & 15.3 & 90 & 79.3 & 1\\
   $M_{t}$ & 44 & 56.4 & 0 & 10.7 & 0 &
   & 66 & 34.3 & 0 & 6.9 & 0 &
   & 78 & 21.7 & 0 & 3.9 & 0\\
        \hline
  &  \multicolumn{5}{c}{{\textbf{Trial} 4}}&&
    \multicolumn{5}{c}{{\textbf{Trial} 5}}&&
    \multicolumn{5}{c|}{{\textbf{Trial} 6}}\\
    \hline
{\textbf{Model}} 
& {$\bar{N}$} & {\textbf{rmse}} & {\textbf{CI}} & {$l_{CI}$} & {\textbf{\%aic}\:\:}  &
& {$\bar{N}$} & {\textbf{rmse}} & {\textbf{CI}} & {$l_{CI}$} & {\textbf{\%aic}\:\:} &
& {$\bar{N}$} & {\textbf{rmse}} & {\textbf{CI}} & {$l_{CI}$} & {\textbf{\%aic}} \\
\hline      
  $M_{z}$  & 205 & 45.9 & 95 & 200.0 & 87 &
  & 199 & 19.8 & 91 & 72.9 & 96 &
  & 201 & 11.3 & 94 & 42.1 & 97\\
   $M_{z_{g_n}}$ & 142 & 62.8 & 45 & 91.5 & 0 &
   & 163 & 39.5 & 19 & 39.9 & 0 &
   & 176 & 25.2 & 14 & 23.9 & 0\\
  $M_{z_f}$  & 93 & 107.4 & 0 & 21.6 & 0 &
  & 133 & 67.1 & 0 & 12.8 & 0 &
  & 159 & 41.6 & 0 & 8.2 & 0\\
   $M_{z_{\tilde{g}_n}}$ & 112 & 90.3 & 7 & 48.5 & 0 &
   & 146 & 54.5 & 0 & 26.0 & 0 &
   & 167 & 33.5 & 1 & 16.9 & 0\\   
  $M_{0}$ & 88 & 112.8 & 0 & 14.9 & 0 &
  & 131 & 68.9 & 0 & 10.9 & 0 &
  & 158 & 42.5 & 0 & 7.0 & 0\\
   $M_{b}$ & 284* & 415.9* & 87 & 3270.3* & 0 &
   & 218* & 130.7* & 93 & 942.1* & 0 &
   & 197 & 18.4 & 94 & 86.8 & 0\\
   $M_{c_1}$ & 136 & 68.4 & 34 & 85.6 & 0 &
   & 165 & 37.3 & 26 & 41.9 & 0 &
   & 180 & 21.6 & 31 & 26.8 & 0\\
   $M_{c_1b}$ & 284* & 415.9* & 87 & 3270.3* & 0 &
   & 218* & 130.7* & 93 & 942.1* & 0 &
   & 197 & 18.4 & 94 & 86.8 & 0\\
   $M_{c_2}$ & 174 & 53.1 & 85 & 192.2 & 11  &
   & 183 & 24.7 & 77 & 62.6 & 1 &
   & 191 & 13.5 & 82 & 36.2 & 2\\
   $M_{c_2b}$ & 284* & 415.9* & 87 & 3270.3* & 2 &
   & 218* & 130.7* & 93 & 942.1* & 3 &
   & 197 & 18.4 & 94 & 86.8 & 1\\
  $M_{t}$ & 87 & 113.0 & 0 & 14.6 & 0 &
  & 131 & 68.9 & 0 & 10.7 & 0 &
  & 158 & 42.5 & 0 & 6.9 & 0\\
  \hline 
\end{tabular}}
\caption{\textit{Simulation study with 100 simulated datasets for each
simulation setting (Trial 1-6) where true generating model
  is $M_z$: empirical average
($\bar{N}$) of the point estimate $\hat{N}$,
root mean square error (rmse), confidence intervals coverage (CI $\%$), 
average length of the confidence intervals ($l_{CI}$) and  percentage
of times each competing model has achieved best AIC (\%aic). 
The $*$ sign denotes the presence of likelihood failure
$\hat{N}=\infty$ \citep{aft:2012}.
We are reporting a finite rmse computed after
removing 
those failure cases. Nominal level of the confidence interval
$1-\alpha=0.95$. }}
\label{tab:5a}
\end{sidewaystable}

\begin{sidewaystable}
\vspace*{15cm}\centering{
\begin{tabular}{|lccccccccccccccccc|}
\hline
 &   \multicolumn{5}{c}{{\textbf{Trial} 7}}& &
    \multicolumn{5}{c}{{\textbf{Trial} 8}}& &
    \multicolumn{5}{c|}{{\textbf{Trial} 9}}\\
    \hline
{\textbf{Model}} 
& {$\bar{N}$} & {\textbf{rmse}} & {\textbf{CI}} & {$l_{CI}$} & {\textbf{\%aic}\:\:}  &
& {$\bar{N}$} & {\textbf{rmse}} & {\textbf{CI}} & {$l_{CI}$} & {\textbf{\%aic}\:\:} &
& {$\bar{N}$} & {\textbf{rmse}} & {\textbf{CI}} & {$l_{CI}$} & {\textbf{\%aic}} \\
\hline
  $M_{z}$ & 91 & 33.3 & 85 & 129.7 & 3 &
  & 87 & 17.3 & 71 & 35.5 & 1 &
  & 90 & 12.0 & 57 & 19.6 & 0\\
   $M_{z_{g_n}}$ & 109.0 & 42.0 & 90 & 170.7 & 93 & 
   & 100 & 13.5 & 92 & 52.8 & 99 &
   & 99 & 7.9 & 93 & 29.3 & 100\\
  $M_{z_f}$ & 49 & 51.3 & 2 & 21.1 & 0 &
  & 70 & 30.3 & 2 & 13.4 & 0 &
  & 82 & 19.2 & 3 & 9.0 & 0\\
   $M_{z_{\tilde{g}_n}}$ & 73 & 36.3 & 60 & 75.9 & 2 &
   & 84 & 18.8 & 60 & 32.8 & 0 &
   & 90 & 12.1 & 59 & 19.7 & 0\\
   $M_{0}$ & 48 & 52.4 & 1 & 17.9 & 1 &
   & 70 & 30.5 & 2 & 13.0 & 0 &
   & 82 & 19.1 & 4 & 8.9 & 0\\
   $M_{b}$ & 209* & 564.0* & 83 & 3866.0* & 0 &
   & 119.4* & 105.4* & 89 & 1776.5* & 0 &
   & 98 & 15.3 & 90 & 193.5 & 0\\
   $M_{c_1}$ & 59 & 42.5 & 29 & 43.2 & 0 &
   & 76 & 25.4 & 11 & 20.8 & 0 &
   & 84 & 16.4 & 16 & 13.1 & 0\\
   $M_{c_1b}$ & 209* & 564.0* & 83 & 3866.0* & 0 &
   & 119.4* & 105.4* & 89 & 1776.5* & 0 &
   & 98 & 15.3 & 90 & 193.5 & 0\\
   $M_{c_2}$ & 69 & 36.4 & 65 & 84.9 & 1 &
   & 80 & 22.4 & 38 & 26.9 & 0 &
   & 86 & 14.8 & 31 & 15.6 & 0\\
   $M_{c_2b}$ & 209* & 564.0* & 83 & 3866.0* & 0 &
   & 119.4* & 105.4* & 89 & 1776.5* & 0 &
   & 98 & 15.3 & 90 & 193.5 & 0\\
  $M_{t}$ & 48 & 52.8 & 1 & 17.4 & 0 &
  & 70 & 30.7 & 2 & 12.8 & 0 &
  & 81 & 19.3 & 3 & 8.7 & 0\\
      \hline 
 &
    \multicolumn{5}{c}{{\textbf{Trial} 10}}& &
    \multicolumn{5}{c}{{\textbf{Trial} 11}}& &
    \multicolumn{5}{c|}{{\textbf{Trial} 12}}\\
    \hline
{\textbf{Model}} 
& {$\bar{N}$} & {\textbf{rmse}} & {\textbf{CI}} & {$l_{CI}$} & {\textbf{\%aic}\:\:}  &
& {$\bar{N}$} & {\textbf{rmse}} & {\textbf{CI}} & {$l_{CI}$} & {\textbf{\%aic}\:\:} &
& {$\bar{N}$} & {\textbf{rmse}} & {\textbf{CI}} & {$l_{CI}$} & {\textbf{\%aic}} \\
\hline
  $M_{z}$  & 170 & 50.2 & 75 & 138.9 & 0 &
  & 173 & 30.8 & 52 & 49.2 & 0 &
  & 182 & 20.6 & 43 & 28.1 & 0\\
   $M_{z_{g_n}}$ & 205 & 46.3 & 91 & 188.1 & 100 &
   & 200 & 18.7 & 95 & 72.7 & 100 &
   & 200 & 11.5 & 92 & 41.8 & 100\\
  $M_{z_f}$ & 96 & 104.3 & 0 & 25.6 & 0 &
  & 139 & 61.2 & 0 & 18.5 & 0 &
  & 164 & 36.6 & 0 & 13.4 & 0\\
   $M_{z_{\tilde{g}_n}}$ & 137 & 68.7 & 34 & 84.5 & 0 &
   & 168 & 34.8 & 35 & 44.8 & 0 &
   & 182 & 20.3 & 45 & 28.4 & 0\\
  $M_{0}$ & 94 & 106.8 & 0 & 21.7 & 0 &
  & 139 & 61.7 & 0 & 17.9 & 0 &
  & 164 & 36.7 & 0 & 13.3 & 0\\
   $M_{b}$ & 284* & 415.9* & 87 & 3270.3* & 0 &
   & 218* & 130.7* & 93 & 942.1* & 0 &
   & 197 & 18.4 & 94 & 86.8 & 0\\
   $M_{c_1}$ & 114 & 87.4 & 3 & 50.2 & 0 &
   & 150 & 50.5 & 1 & 28.7 & 0 &
   & 170 & 30.7 & 2 & 18.9 & 0\\
   $M_{c_1b}$ & 284* & 415.9* & 87 & 3270.3* & 0 &
   & 218* & 130.7* & 93 & 942.1* & 0 &
   & 197 & 18.4 & 94 & 86.8 & 0\\
   $M_{c_2}$ & 130 & 73.6 & 31 & 85.2 & 0 &
   & 158 & 43.9 & 8 & 36.5 & 0 &
   & 174 & 27.2 & 8 & 22.2 & 0\\
   $M_{c_2b}$ & 284* & 415.9* & 87 & 3270.3* & 0 &
   & 218* & 130.7* & 93 & 942.1* & 0 &
   & 197 & 18.4 & 94 & 86.8 & 0\\
  $M_{t}$ & 93 & 107.2 & 0 & 21.3 & 0 &
  & 139 & 61.9 & 0 & 17.7 & 0 &
  & 164 & 36.8 & 0 & 13.2 & 0\\      
      \hline
\end{tabular}}
\caption{\textit{Simulation study with 100 simulated datasets for each
simulation setting (Trial 7-12) where true generating model
  is $M_{z_{g_n}}$: empirical average
($\bar{N}$) of the point estimate $\hat{N}$,
root mean square error (rmse), confidence intervals coverage (CI $\%$), 
average length of the confidence intervals ($l_{CI}$) and  percentage
of times each competing model has achieved best AIC (\%aic). 
The $*$ sign denotes the presence of likelihood failure
$\hat{N}=\infty$ \citep{aft:2012}.
We are reporting a finite rmse computed after
removing 
those failure cases. Nominal level of the confidence interval
$1-\alpha=0.95$. }}
\label{tab:5b}
\end{sidewaystable}

\section{Concluding remarks and discussion}
\label{sec:close}

In order to model behavioural effect to capture and 
other possible longitudinal patterns
we have proposed a flexible model framework based
on the conditional probability parameterization and 
a suitable ordering and scaling of the binary sequences 
representing the individual partial capture histories. 
One meaningful ordering is built up through the 
binary representation of integers
corresponding to each conditioning sequence of partial capture history. 
Then, the integer quantity 
representing the numerical quantification 
of a partial capture history 
is appropriately rescaled
in order to obtain  a suitable quantitative covariate 
$z$ ranging in a standard interval $[0,1]$.
We have provided a natural interpretation of such a 
covariate $z$ 
as a meaningful proxy for a \textit{memory effect} and 
formal correspondence with the Markovian dependence.
We have also discussed
some other alternative quantifications. 
The basic idea of the new model framework
can be easily implemented 
within 
the setup of 
a logistic model where 
each capture occurrence $x_{ij}$ is considered as a binary outcome 
with $r(z_{ij}$) 
as the basic linear predictor of the log-odds of the corresponding
probability. 
The function $r$ could be either a linear or non linear function of 
$z_{ij}$. 
In this case, when the non linear function is a step function
it turns out to 
partition conditional probability
parameters into equivalence classes
possibly recovering known standard behavioural 
or temporal 
models such as $M_{b}$ or $M_{c_k}$ and $M_{t}$
and discovering new meaningful ones
such as $M_{z.cut(k)}$.


Indeed the use of this general framework allowed us to revisit some
well known datasets and discover new parsimonious behavioural patterns 
that better fit the the observed data. 
The discernibility of new non-enduring patterns with
respect to already available 
enduring or ephemeral behavioral effects has been verified with a
simulation study where the AIC criterion is able to recover the
new pattern in most simulated datasets.
Point and interval estimates yield convincing results in terms of
small RMSE and  adequate coverage.

Unconditional likelihood inference is easily implemented  
recycling consolidated standard GLM
routines.
An integrated suite of R 
\citep{R:2013}
functions have been developed and are
available as an R package upon request.

We hint also at a possible extension outside the closed
capture-recapture context of the quantization idea.
The same idea can be applied 
more generally to model memory effects in studies with 
longitudinal binary outcomes where 
binary events such as successful surgery experiences 
or correctly performed tasks are observed. 
Also it is possible to generalize this strategy to categorical-ordinal data 
using an appropriate scaling.

There are certainly other issues which should be addressed for 
a more thorough understanding of real data such as 
allowing for heterogeneous capture probabilities and including  
the possible presence of individual covariates. While the former aspect
can be easily accommodated and implemented as already argued 
within the standard GLM 
framework through the addition of a longitudinal individual random effect
the latter is more difficult to be embedded in the proposed 
inferential setting which uses the unconditional likelihood. In this
case individual covariates would not be available for unobserved units. 
Indeed possible alternative ways out are the use of conditional
likelihood
or implementing our models 
using data augmentation
within a Bayesian framework following the approach in 
\cite{royl:2009}. We actually plan to develop this in a future work. 
Actually some previous work on 
alternative inferential approaches for standard behavioural models
\cite{aft:2012} suggests that the Bayesian approach could be more 
promising.
Here we have focussed more specifically on understanding the role, 
meaning and possible alternative uses
of the new memory-effect covariates and their connections with 
already available models.



\bibliographystyle{abbrvnamed}

\bibliography{biblio}

\newpage

\begin{center}
{\LARGE \bf Supplementary Web Materials for\\ 
``Flexible behavioral capture-recapture modelling''
}
\end{center}

\setcounter{section}{0}
\renewcommand{\thesection}{S\arabic{section}}

\section{Further examples of meaningful models based on partitions}

We have seen how some classical models 
($M_b$, $M_{c_{1}}$ and $M_{c_{2}}$) 
correspond to 
specific ways of
partitioning
the set of all partial capture histories 
$H = \cup_{b=1}^B H_{b}$ 
and
setting equal all the conditional probabilities
with conditioning partial capture history $\mathbf{x}$
belonging to the same partition set $H_{b}$ as follows 
\begin{eqnarray*}
p_{l_{\mathbf x}+1}(\mathbf{x})=p_{H_{b}} 
\qquad
\forall \quad \mathbf{x} \in H_{b} 
\end{eqnarray*}
Now we provide other instances of partitions 
which can be associated to different classical 
and new models accounting for longitudinal 
behavioral and temporal patterns.

We  will show four partitions
corresponding to the so-called
{\em time-effect} model $M_t$, the mixed {\em ephemeral-enduring} effect model $M_{c_{1}b}$
introduced in \cite{Yang:Chao:2005}, the behavioural {\em
  vanishing-effect} model $M_{L_2}$ proposed in \cite{Farc:2011}
and, as a new proposal, an alternative behavioural 
model
denoted with $M_{count}$ 
where the capture probabilities vary according 
to the {\em absolute} number of previous captures occurred.
To simplify notation and understanding let us consider 
a discrete capture-recapture experiment with $t=5$ 
capture occasions. We start with the classical 
{\em time-effect} model $M_t$ which 
corresponds to the following partition in $t=5$ subsets
\begin{eqnarray*}
\mathcal{H}_{5}(M_{t})\:\: :
\begin{cases}
H_1=\{ () \} \qquad\qquad\qquad\qquad\qquad\\
H_2=\{ (0),(1) \}  \\
H_3=\{ (0,0),(1,0),(0,1),(1,1) \}  \\
H_4=\{ (0,0,0),(1,0,0),(0,1,0),(1,1,0),(0,0,1),(1,0,1),(0,1,1),(1,1,1) \} \\
H_5=\{ (0,0,0,0),(1,0,0,0),(0,1,0,0),(1,1,0,0),(0,0,1,0),(1,0,1,0),\\
\qquad \quad(0,1,1,0),(1,1,1,0),(0,0,0,1),(1,0,0,1),(0,1,0,1),\\ 
\qquad \quad(1,1,0,1),(0,0,1,1),(1,0,1,1),(0,1,1,1),(1,1,1,1)\}
\end{cases}	
\end{eqnarray*}
The set $H$ is partitioned 
according just to the length of partial capture histories
which identifies each capture occasion
without considering the pattern of the occurrences. 
In fact with this partition we are modelling
a temporal pattern rather than 
a behavioural effect.

The mixed {\em ephemeral-enduring} effect model $M_{c_{1}b}$
introduced in \cite{Yang:Chao:2005} can be regarded as a 
model where conditional probabilities are grouped according 
to conditioning partial capture histories belonging to subsets 
of the following partition:
\begin{eqnarray*}
\mathcal{H}_{3}(M_{c_{1}b}):
\begin{cases}
H_{1}=\{ (),(0),(0,0),(0,0,0),(0,0,0,0)\\
H_{2}=\{ (1,0),(1,0,0),(0,1,0),(1,1,0),\\
\qquad \quad(1,0,0,0),(0,1,0,0),(1,1,0,0),(0,0,1,0),(1,0,1,0),(0,1,1,0),(1,1,1,0) \}\\
H_{3}= \{  (1), (0,1),(1,1),(0,0,1),(1,0,1),(0,1,1),(1,1,1),(0,0,0,1),\\ 
\qquad \quad (1,0,0,1),(0,1,0,1),(1,1,0,1),(0,0,1,1),(1,0,1,1),(0,1,1,1),(1,1,1,1) \} 
\end{cases}	
\end{eqnarray*}
As in model $M_{c_{1}}$ we partition the set $H$ according to
the occurrence (0 or 1) in the last position. 
However, differently from the standard first order Markovian model,
in correspondence of 
the same conditioning event $x_{l_{\mathbf{x}}}=0$,
one distinguishes  
those histories $\mathbf{x}$ where a previous capture has occurred at least once
($H_2$) from those where no previous capture has occurred ($H_1$).

As defined in \cite{Farc:2011} ``model $M_{L_2}$ corresponds to a vanishing
behavioural effect if the animal is not captured in most recent occasion, 
or captured only once in the last three occasions''. It corresponds to
the bipartition
\begin{eqnarray*}
\mathcal{H}_{2}(M_{L_2})\:\: :
\begin{cases}
H_{1}=\{(), (0), (1,0), (0,0,0)(1,0,0),(0,1,0),(1,1,0), (0,0,1),\\
\qquad \quad(0,0,0,0)(0,1,0,0),(0,0,1,0),(0,1,1,0), (0,0,0,1),\\
\qquad \quad (1,0,0,0)(1,1,0,0),(1,0,1,0),(1,1,1,0), (1,0,0,1)\}\\
H_{2}=\{(1),(0,1),(1,1),(1,0,1),(0,1,1),(1,1,1),(0,1,0,1),\\
\qquad \quad (0,0,1,1),(0,1,1,1),(1,1,0,1),(1,0,1,1),(1,1,1,1)\}
\end{cases}	
\end{eqnarray*}
Indeed, it can be regarded also as a specific constrained 
Markovian model of order 3.

Another model which can be of some interest is expressed in terms of the following partition
\begin{eqnarray*}
\mathcal{H}_{5}(M_{count})\:\: :
\begin{cases}
H_1=\{ (),(0),(0,0), (0,0,0), (0,0,0,0)\}\\
H_2=\{ (1) (1,0),(0,1),(1,0,0),(0,1,0),(0,0,1),\\
\qquad \quad (1,0,0,0),(0,1,0,0),(0,0,1,0),\},(0,0,0,1)  \\
H_3=\{ (1,1),(1,1,0),(1,0,1),(0,1,1),(1,1,0,0),\\
\qquad \quad(1,0,1,0),(0,1,1,0), (1,0,0,1),(0,1,0,1), ,(0,0,1,1)\}  \\
H_4=\{ (1,1,1),(1,1,1,0),(1,1,0,1),(1,0,1,1),(0,1,1,1) \} \\
H_5=\{ (1,1,1,1)\} 
\end{cases}	
\end{eqnarray*}
In each subset of the partition $\mathcal{H}_{5}(M_{count})$ 
the partial capture histories have the same number of captures (the same number of 1's).
Notice also that the partition 
$\mathcal{H}_{5}(M_{count})$
shares the same subset $H_1$ of 
$\mathcal{H}_{3}(M_{c_{1}b})$  and 
$\mathcal{H}_{2}(M_{b})$.
In view of the likelihood factorization already argued in 
\cite{aft:2012} we have 
\begin{eqnarray*}
&L(N,\mathbf{p}_{\mathcal{H}_{B}}) \propto 
\left[
{{N}\choose{M}} p_{H_{1}}^{n_{(H_{1}1)}}(1-p_{H_{1}})^{n_{(H_{1}0)}+t(N-M)}
\right]
\prod_{b=2}^{B} p_{H_{b}}^{n_{(H_{b}1)}}(1-p_{H_{b}})^{n_{(H_{b}0)}}  
\end{eqnarray*}
where, for each $b=1,2,...,B$, $n_{(H_{b}0)}$ is the number of times that all the observed units which experience partial capture history $\mathbf{x}$ belonging to $H_{b}$ are not captured at time $l_{\mathbf{x}}+1$; similarly $n_{(H_{b}1)}$ is the number of times that the observed units  which  experience partial capture history $\mathbf{x}$ belonging to $H_{b}$ are captured at time $l_{\mathbf{x}}+1$. Formally $\forall \: b=1,\dots ,B$ 
\begin{eqnarray*}
& n_{(H_{b}0)}=\sum_{i=1}^{M}\sum_{\mathbf{x} \in H_{b}}I\left[ (x_{i1},\dots ,x_{il_{\mathbf{x}}})= \mathbf{x} \: , \: x_{il_{\mathbf{x}}+1}=0\right]  \\
& n_{(H_{b}1)}=\sum_{i=1}^{M}\sum_{\mathbf{x} \in H_{b}}I\left[ (x_{i1},\dots ,x_{il_{\mathbf{x}}})= \mathbf{x} \: , \: x_{il_{\mathbf{x}}+1}=1\right] 
\end{eqnarray*}
This implies that all models sharing the same counts
$n_{(H_{1}0)}$ and $n_{(H_{1}1)}$ 
(and this is always true when they share the same $H_1$)
have the same profile likelihood for $N$
and yield  the same point and interval estimates for $N$.
Hence this is true for models $M_{count}$, 
$M_{c_{1}b}$ and $M_b$.

\section{Proof of the relationship of the numerical quantification $g(\mathbf{x})$
 with Markovian models of arbitrary order $k$}
 
In order to prove the relationship between the proposed 
numerical quantification 
$z=g(\mathbf{x})$
and the generic $k$-th order Markovian structure expressed 
through the conditional probabilities 
$p_{l_{\mathbf{x}}+1}(\mathbf{x})$
we will  show 
a correspondence beween the range 
of $z=g(\mathbf{x})$
and the range 
$\mathcal{X}^{k}$ 
of the last $k$ digits
of $\mathbf{x}$
such that
$$
z = g(\mathbf{x}) \in I_r 
\Longleftrightarrow
\sum_{p=1}^{k}x_{l_{\mathbf{x}}-k+p}\, 2^{p-1} 
= r-1
$$
where $I_{1}=[0,1/2^{k}]$ and 
$I_{r} =\left( \frac{r-1}{2^k},\frac{r}{2^k}\right]$
for any $r=2,...,2^k$. In fact, this 
means that 
the subset  $I_r$ which
$g(\mathbf{x})$
turns out to belong to  
depends only on the last $k$
digits of $\mathbf{x}$.
Hence 
if $s(z)$ is a step function defined as follows 
\begin{eqnarray*}
s(z)=logit(p_{H_{r}}) \qquad \forall z \in I_{r}  
\end{eqnarray*}
for any $r=1,...,2^k$,
the conditional probability  corresponding to 
our
logistic framework based on $z$
$$
Pr \left(
X_{i {l_{\mathbf{x}}+1}} = 1 \left|
X_{i {1}} = x_{1},
..., 
X_{i {l_{\mathbf{x}}-k+1}} = x_{l_{\mathbf{x}}-k+1},...,
X_{i {l_{\mathbf{x}}}} = x_{l_{\mathbf{x}}}
\right. 
\right)
= s(g(\mathbf{x})) = s(z)
$$
depends on 
the binary configuration of 
the last $k$ binary digits (occurrences)
as prescribed in 
any $k$-th order Markovian model
\begin{eqnarray*}
Pr \left(
X_{i {l_{\mathbf{x}}+1}} = 1 \left|
X_{i {1}} = x_{1},
..., 
X_{i {l_{\mathbf{x}}-k+1}} = x_{l_{\mathbf{x}}-k+1},...,
X_{i {l_{\mathbf{x}}}} = x_{l_{\mathbf{x}}}
\right. 
\right)=\\
= Pr \left(X_{i {l_{\mathbf{x}}+1}} = 1 \left|
X_{i {l_{\mathbf{x}} -k+1}} = x_{l_{\mathbf{x}}-k+1},
..., 
X_{i {l_{\mathbf{x}}}} = x_{l_{\mathbf{x}}} 
\right.  \right) .
\end{eqnarray*}
Indeed 
for any $r \in \{1,...,2^k\}$
there is only one configuration 
$(x_{l_{\mathbf{x}}-k+1},...,x_{l_{\mathbf{x}}}) \in \mathcal{X}^{k}$
such that 
$$
\sum_{p=1}^{k}x_{l_{\mathbf{x}}-k+p}\, 2^{p-1} = 
r -1.
$$
In order to fully understand the 
ensuing behavioural 
model based on the partition 
induced by the partitioning 
of the 
covariate range $[0,1]=\cup_{r=1}^{2^{k}} I_r$, 
we need to  distinguish
 two cases: 
\begin{itemize}
\item  the case where the conditioning 
partial capture history $\mathbf{x}$ 
is a binary sequence with length 
greater than or equal to $k$ ($l_{\mathbf{x}}\geq k$) 
\item 
the case where the conditioning binary sequence 
$\mathbf{x}$ 
has length less than $k$ ($l_{\mathbf{x}}< k$).
\end{itemize}
The latter case does not involve
the $k$-th order 
Markov property 
but it affects 
the parameterization related to the initial conditional probabilities 
which can be 
defined in a somewhat arbitrary fashion. 
We will look at each case separately in the following two subsections.

\subsection{Mapping $z=g(\mathbf{x})$ with partial histories with length greater than or equal to $k$}

Recall that for any fixed positive integer $k$ there is a one-to-one 
mapping between the possible configurations of $k$ digits 
$\mathcal{X}^{k}$ and the first $2^k$ non negative integers 
$\{0,1,...,2^{k}-1\}$.
Let us consider a generic partial capture history 
$\mathbf{x}$ of length $l_{\mathbf{x}}\geq k$. 

In the definition of the basic mapping 
\begin{eqnarray*}
f(\mathbf{x}) = f(x_{1},\dots ,x_{l_{\mathbf{x}}}) = \sum_{j=1}^{l_{\mathbf{x}}} x_{j} 2^{j-1} \in \{0,1,2,...,2^{l_{\mathbf{x}}}-1\}
\end{eqnarray*}
used to build
up our 
$z=g(\mathbf{x})$
the sum involving the last $k$ 
digits can take $2^k$ different values. 
Formally it can be written as 
$$
\sum_{j=l_{\mathbf{x}}-k+1}^{l_{\mathbf{x}}}x_{j}\, 2^{j-1} \: =\: 2^{l_{\mathbf{x}}-k}\sum_{p=1}^{k}x_{l_{\mathbf{x}}-k+p}\, 2^{p-1}
$$
which for $(x_{l_{\mathbf{x}}-k+1},\dots x_{l_{\mathbf{x}}}) \in \mathcal{X}^{k}$
takes value in $\{c \cdot 0,c \cdot 1,\dots ,c \cdot y\}$
where $c=2^{l_{\mathbf{x}}-k}$.
Hence
the ratio which defines the function $g$ can be rewritten as follows
\begin{align}
\frac{\sum_{j=l_{\mathbf{x}}-k+1}^{l_{\mathbf{x}}}x_{j}\, 2^{j-1}}{2^{l_{\mathbf{x}}}-1} \: =\:
\frac{\sum_{p=1}^{k}x_{l_{\mathbf{x}}-k+p} \, 2^{p-1}}{2^{k}-\frac{1}{2^{l_{\mathbf{x}}-k}}} 
\: \in \: \left\lbrace 0,\frac{1}{2^{k}-\frac{1}{2^{l_{\mathbf{x}}-k}}}, \dots ,\frac{2^{k}-1}{2^{k}-\frac{1}{2^{l_{\mathbf{x}}-k}}}\right\rbrace 
\label{mckappendix}
\end{align}
On the other hand, considering the sum involving the first
$(l_{\mathbf{x}}-k)$ digits 
of the binary sequence
we can get the following bounds
\begin{align}
0\:\leq\: \frac{ \sum_{j=1}^{l_{\mathbf{x}}-k}x_{j} \,2^{j-1}}{2^{l_{\mathbf{x}}}-1} 
\:\leq\: \frac{2^{l_{\mathbf{x}}-k}-1}{2^{l_{\mathbf{x}}}-1}
\: =\:\frac{1-\frac{1}{2^{l_{\mathbf{x}}-k}}}{2^{k}-\frac{1}{2^{l_{\mathbf{x}}-k}}}
\label{mcc}
\end{align}
For any $(x_{l_{\mathbf{x}}-k+1},\dots , x_{l_{\mathbf{x}}}) \in
\mathcal{X}^{k}$ 
from \eqref{mckappendix} we can represent 
\begin{align}
\frac{\sum_{j=l_{\mathbf{x}}-k+1}^{l_{\mathbf{x}}}x_{j}\, 2^{j-1}}{2^{l_{\mathbf{x}}}-1}=\frac{r-1}{2^{k}-\frac{1}{2^{l_{\mathbf{x}}-k}}} 
\label{BBB}
\end{align}
for some $r\in \{1, \dots ,2^{k}\}$
so that the following inequalities hold 
\begin{align}
\frac{r-1}{2^{k}} \: < \: \frac{r-1}{2^{k}-\frac{1}{2^{l_{\mathbf{x}}-k}}}
\: \: \leq \: \: 
\frac{\sum_{j=1}^{l_{\mathbf{x}}}x_{j}\,
  2^{j-1}}{2^{l_{\mathbf{x}}}-1}  
= g(\mathbf{x}) 
\: \: \leq \:\:
 \frac{1-\frac{1}{2^{l_{\mathbf{x}}-k}}}{2^{k}-\frac{1}{2^{l_{\mathbf{x}}-k}}}
 + \frac{r-1}{2^{k}-\frac{1}{2^{l_{\mathbf{x}}-k}}} 
 \: \leq \: \frac{r}{2^{k}}
\label{AAA}
\end{align}
Indeed the second inequality follows from the fact that the {\em rhs}  has the sum running
over all the elements of the binary sequence while the {\em lhs}  corresponds to \eqref{BBB} where 
the sum runs over the last $k$ elements only. The third inequality follows 
combining 
\eqref{mcc} and \eqref{BBB}. Finally, the last inequality follows
from the fact that 
$\forall \: r \: \in \: \{1, \dots ,2^{k}\}$ we have
\begin{align*}
\frac{r-\frac{1}{2^{l_{\mathbf{x}}-k}}}{2^{k}-\frac{1}{2^{l_{\mathbf{x}}-k}}}-\frac{r}{2^{k}}=
\frac{2^{l_{\mathbf{x}}-k}\left[r-\frac{1}{2^{l_{\mathbf{x}}-k}} \right]}{2^{l_{\mathbf{x}}}-1}-\frac{r}{2^{k}}=
\frac{r-2^{k}}{(2^{l_{\mathbf{x}}}-1)2^{k}}\leq 0
\end{align*}
In this way we have formally proved that for any 
partial capture history $\mathbf{x}^{'}$ and $\mathbf{x}^{''}$ 
sharing the same last $k$ digits we have that $g(\mathbf{x}^{'})\in I_{r}$ and $g(\mathbf{x}^{''})\in I_{r}$
for a suitable integer 
$$
r-1= \sum_{p=1}^{k}x_{l_{\mathbf{x}}-k+p}\, 2^{p-1}.
$$
This implies that $z =g(\mathbf{x}) \in I_{r}$ 
if and only if 
the binary configurations of last $k$ digits 
of $\mathbf{x}$ correspond to the integer $r-1$. 
These $k$ digits in fact correspond to the last $k$ occurrences
of each partial capture history in $H_{r}$
provided there are at least as much. 
This essentially leads us to a 
Markovian model of order $k$.

 \subsection{Mapping  $z=g(\mathbf{x})$ with partial histories with length less than $k$}

We begin with a simple example 
of Markov model of order $k=2$ in
a capture-recapture experiment with $t=5$ occasions.
It points out some critical 
aspects in recovering the desired Markovian 
model  
when using 
the covariate $z=g(\mathbf{x})$ also for partial capture histories 
$\mathbf{x}$ with length $l_{\mathbf{x}}<k$.
In fact, if we 
look at the numerical values of $z=g(\mathbf{x})$
displayed in Table~\ref{tab:x} of our main paper and 
we keep on  partitioning the set $H$ 
according to the value of $g(\mathbf{x})$ into the subintervals 
$I_1 =\left[ 0,\frac{1}{4}\right]$, 
$I_{2} =\left[ \frac{1}{4},\frac{2}{4}\right]$,
$I_{3} =\left[ \frac{2}{4},\frac{3}{4}\right]$,
$I_{4} =\left[ \frac{3}{4},1\right]$
we obtain the following partition of the set $H$
\begin{eqnarray}
\label{modelmc2}
\mathcal{H}_{4}(M_{*})=
\begin{cases}
H_1=\{ (),(0),(00),(000),(100),(0000),(0100),(1000),(1100) \} \\
H_2=\{ (10),(010),(110),(0010),(0110),(1010),(1110) \}  \\
H_3=\{ (01),(001),(101),(0001),(0101),(1001),(1101) \}  \\
H_4=\{ (1),(11),(011),(111),(0011),(0111),(1011),(1111) \}  
\end{cases}	
\end{eqnarray}
Differently from the partition 
corresponding to model $M_{c_{2}}$ 
considered in 
\cite{Farc:2011} and 
denoted as 
$\mathcal{H}_{4}(M_{c_2})$
in formula (4) 
of our main article 
the partial capture history $(1)$, 
corresponding to one capture in the first occasion, 
belongs to the subset 
$H_{4}$ instead of $H_{3}$. 
In fact, 
looking at the partial histories with 
$l_{\mathbf{x}}\leq 1$
we have that 
$\mathbf{x}'=()$ and $\mathbf{x}''= (0)$ are mapped in 
$z'=g(\mathbf{x}')=0$ and 
$z''=g(\mathbf{x}'')=0$ and they both fall
in
the same interval $I_1$  
where also $(0,0)$ is mapped. On the other hand 
the partial capture history $\mathbf{x}=(1)$ is mapped 
in $z=g(\mathbf{x})=1$  which falls in 
the interval $I_4$ where also
$(1,1)$ is mapped. 
This last mapping somehow breaks the correspondence with 
$\mathcal{H}_{4}(M_{c_{2}})$ as in 
in formula (4) 
of our main article 
although the model associated to 
$\mathcal{H}_{4}(M_{*})$
is still Markovian of order 2
since the initial conditional probabilities are 
irrelevant for the Markov property.
The difference between the two partitions 
 depends on the 
arbitrary ways in which 
one can define 
the initial conditional 
probabilities 
namely those whose conditioning events
correspond to 
partial capture histories 
whose length is less than $k$. 
If one likes to recover exactly the partition 
$\mathcal{H}_{4}(M_{c_2})$
by means of partitioning the covariate range
there is  a simple modification to fix that.
For a generic partial capture history $\mathbf{x}$ with 
length $l_{\mathbf{x}}<k$
obviously one cannot  get 
the usual dependence on the last $k$ digits
since the length of $\mathbf{x}$ is smaller.
In this case one may artificially complete 
the partial capture 
history augmenting it 
in a conventional 
way with $k-l_{\mathbf{x}}$ fictitious digits ahead. 
In this way, we are back to 
dealing with a partial capture history  with at least $k$ digits
as in the previous subsection.
Indeed in the Markovian models $M_{c_{k}}$ proposed in 
\cite{Yang:Chao:2005} and \cite{Farc:2011}
it is assumed that the $k-l_{\mathbf{x}}$  
unobserved/missing/imaginary previous digits
are 
all set equal to 0. 

In our previous example if we insert $k=2$ zeroes ahead
of each actually observed partial capture history 
$\mathbf{x} =(x_{1},\dots , x_{l_{\mathbf{x}}})$ 
and denote the augmented sequence with $\mathbf{x}_{aug}$
we can then 
basically recover the partition $\mathcal{H}_4(M_{c_{2}})$
as follows 
\begin{eqnarray*}
\begin{cases}
H_1  =  \{  
(\underline{0,0}),(\underline{0,0},0),(\underline{0,0},0,0),(\underline{0,0},0,0,0),(\underline{0,0},1,0,0),
(\underline{0,0},0,0,0,0),\\
\, \,  \, \,  \, \,  \, \, \, \,  \, \,  \, \,  \, \,  \,  \, \,   (\underline{0,0},0,1,0,0),(\underline{0,0},1,0,0,0),(\underline{0,0},1,1,0,0) \} \\
H_2  = \{
(\underline{0,0},1,0),(\underline{0,0},0,1,0),(\underline{0,0},1,1,0),(\underline{0,0},0,0,1,0),\\
\, \,  \, \,  \, \,  \, \, \, \,  \, \,  \, \,  \, \,  \,  \, \,   (\underline{0,0},0,1,1,0),(\underline{0,0},1,0,1,0),(\underline{0,0},1,1,1,0)
\},\\ 
H_3  =  \{
(\underline{0,0},1),(\underline{0,0},0,1),(\underline{0,0},0,0,1),(\underline{0,0},1,0,1),\\
\, \,  \, \,  \, \,  \, \, \, \,  \, \,  \, \,  \, \,  \,  \, \,    (\underline{0,0},0,0,0,1),(\underline{0,0},0,1,0,1),(\underline{0,0},1,0,0,1),(\underline{0,0},1,1,0,1) \}  \\
H_4  =  \{
(\underline{0,0},1,1),(\underline{0,0},0,1,1),(\underline{0,0},1,1,1),\\
\, \,  \, \,  \, \,  \, \, \, \,  \, \,  \, \,  \, \,  \,  \, \,   (\underline{0,0},0,0,1,1),(\underline{0,0},0,1,1,1),(\underline{0,0},1,0,1,1),(\underline{0,0},1,1,1,1) \} 
\end{cases}	
\end{eqnarray*}

\noindent which yields the correspondence $\mathbf{x}_{aug} \in H_{r}$ 
if and only if 
$g(\mathbf{x}_{aug}) \in I_{r}$.
Notice that we have marked the imputed initial segment with an underlying sign.
In this way the empty partial capture history $()$ changes in $(\underline{0,0})$,
$(0)$ changes in $(\underline{0,0},0)$, 
$(1)$ changes in $(\underline{0,0},1)$ and so on.

More generally and formally in order to 
use our approach of 
partitioning $H$ through  subintervals 
$I_1$,...,$I_{2^k}$
of  a suitable numerical covariate
to get the correspondence 
with
the $k$-th order 
Markovian models 
as in
\cite{Yang:Chao:2005} and  
\cite{Farc:2011} 
one  can modify 
the definition of the original numeric summary 
$z=g(\mathbf{x})$
slightly changing its argument $\mathbf{x}$ into an augmented 
history $\mathbf{x}_{aug} = (0,..,0,\mathbf{x})$
with $k-l_{\mathbf{x}}$ zeroes ahead as follows
$$
g_{aug}(\mathbf{x}) = g(\mathbf{x}_{aug}).
$$
This can be formalized in matrix notation  considering 
the whole binary capture history matrix  $\mathbf{X}$
and 
deriving an augmented matrix
 $\mathbf{X}_{aug}=[\underline{\mathbf{0,...,0}},\mathbf{X}]$
obtained by adding $k$ columns of zeros on the left side of the 
original matrix $\mathbf{X}$. 
One can compute in the usual way
the corresponding covariate matrix $\mathbf{Z}_{aug}$
by applying the original function $g$ to all partial capture 
histories in $\mathbf{X}_{aug}$.
At this point,
instead of the former matrix $\mathbf{Z}$ built directly from $\mathbf{X}$
one uses as covariate matrix 
only the last $t$ columns of  $\mathbf{Z}_{aug}$.

\section{Models driven by partitions of the range of meaningful behavioural covariates and the search of
optimal partitions}

Let us now show that model $M_{L_2}$ proposed in \cite{Farc:2011}
can be recovered
within our general logistic regression framework
which relies 
on the meaningful numeric covariate $z$ adopting 
as a regression function a step function
with only two levels corresponding to the bipartition of the 
range of $z$ into two contiguous intervals: $[0,0.625];(0.625,1]$.  
In fact, for the initial partial capture histories $(1)$, $(01)$ and
for all 
other $\mathbf{x}$ with $l_{\mathbf{x}}\geq 3$ such that
$(x_{1},\dots ,x_{l_{\mathbf{x}-3}},1,0,1)$, $(x_{1},\dots ,x_{l_{\mathbf{x}-3}},0,1,1)$ and 
$(x_{1},\dots ,x_{l_{\mathbf{x}-3}},1,1,1)$ we have that $z>0.625$.
This can be easily checked numerically for the first two 
histories $\mathbf{x}_1=(1)$ and $\mathbf{x}_2= (01)$ 
since $g(\mathbf{x}_1)=1$ and  $g(\mathbf{x}_2)=2/3$.
For all the other partial capture histories
we can  focus 
on the last three digits. We have already argued that  there are $8=2^3$
subintervals 
$I_{1}=[0,1/2^{3}]$ and 
$I_{r} =\left( \frac{r-1}{2^3},\frac{r}{2^3}\right]$
for any $r=1,...,2^3$
such that 
$z \in I_r$ if and only if $\sum_{p=1}^{3}x_{l_{\mathbf{x}}-3+p}\, 2^{p-1} 
= r-1$. Since $(0.625,1]=I_6 \cup I_7 \cup I_8$ one can easily verify 
numerically that the only three last digits 
$(x_{l_{\mathbf{x}-2}},x_{l_{\mathbf{x}-1}},x_{l_{\mathbf{x}}})$ such
that 
$\sum_{p=1}^{3}x_{l_{\mathbf{x}}-3+p}\, 2^{p-1}  = r-1\geq 5$ are 
$(1,0,1)$, $(0,1,1)$ and $(1,1,1)$.
Hence the
intervals 
$[0,0.625]$ and $(0.625,1]$
lead to the same bipartition of the set $H$
considered in model $M_{L_2}$.
We also remark  that in light of our argument
there is an underlying  correspondence 
between model $M_{L_2}$ and 
Markovian models of order $k=3$ related to 
partition intervals $I_{r} =\left(
  \frac{r-1}{2^3},\frac{r}{2^3}\right]$
of the behavioural covariate 
$z=g(\mathbf{x})$. In fact  we can regard model $M_{L_2}$ as a 
simplified reduced Markovian model of order $k=3$.

Indeed once acknowledged that model $M_{L_2}$ corresponds to one of the
possible bipartition of the range of $z$ one can wonder whether there
are other bipartitions which can fit the data better.
This naturally leads us to look for an optimal bipartition  
of the range 
in terms of the AIC
resulting from the corresponding model.
We considered models associated to 
alternative intervals $[0,e_1] \cup (e_1,1]$
and eventually determine the best cutpoint 
$e_1^* \in [0,1]$
denoting the corresponding 
bipartition  of partial capture histories
with 
$\mathcal{H}_{2}(M_{z.cut(1)}) = \{H_1,H_2\}$  
where $H_1$ and $H_2$  are such that 
$$
\mathbf{x}\in H_{1} 
\Longleftrightarrow
g(\mathbf{x}) \in [0,e_1^*]  
\qquad ; \qquad 
\mathbf{x}\in H_{2} 
\Longleftrightarrow
g(\mathbf{x}) \in (e_1^*,1]  
$$
In fact, in our applications we found  
the optimal cutpoint  through a simple finite grid search
among all the values
$e_1=g(\mathbf{x})$ corresponding to an actually observed 
partial capture history $\mathbf{x}$.
We remark that the optimal single cut found 
in the Great Copper example ends up being 
$e_1^*=0.625$ which actually corresponds to model 
$M_{L_2}$. 

Moreover, this idea can be extended to more than one cutpoint. 
In 
this case the computational burden for the finite grid search 
with cutpoints corresponding to actually observed 
partial capture histories becomes heavier. Despite that 
in all our real data applications and simulations
we were able to
easily implement the full search up to two cutpoints. 
For more than two cutpoints 
we considered two alternative strategies:
performing a simplified search
reducing the set of possible cuts to a subset of quantifications
of actually observed 
partial capture histories or 
starting from previously determined optimal cuts
and looking for a further cut which 
is located in between.

\end{document}